\documentclass[sigconf,balance=false,nonacm]{acmart} %
\usepackage{xurl} 
\usepackage{longtable}
\usepackage{hyperref}
\usepackage{pdflscape} 
\usepackage[normalem]{ulem}

\newcommand{\autorefappendix}[1]{\hyperref[#1]{\ref*{#1}}}  

\newcommand*\grcell{\cellcolor{gray!25}}

\usepackage[skip=0pt]{caption} 
\usepackage{graphicx} 
\usepackage[table]{xcolor} 
\definecolor{mygray}{rgb}{0.83, 0.83, 0.83}
\newcommand{\cut}[1]{}

\usepackage{paralist}
\usepackage{multicol}
\usepackage{multirow}
\usepackage{makecell}
\usepackage{csquotes}
\usepackage{fontspec} 

\usepackage{tablefootnote}
\usepackage{cleveref}
\newfontfamily\arabicfont[
    Script=Arabic,
    Scale=MatchLowercase
]{Amiri-Regular.ttf} 

\newcommand{\ar}[1]{{\arabicfont #1}}
\DeclareSymbolFont{cmsysymbols}{OMS}{cmsy}{m}{n}
\DeclareMathSymbol{\mytimes}{\mathbin}{cmsysymbols}{"02}

\usepackage{siunitx} 
\usepackage{makecell}
\usepackage{threeparttable}
\usepackage{tabularx}
\usepackage{tcolorbox}
\setcopyright{cc}
\setcctype{by}

\begin{document}

\title[]{Evaluating PDPL Compliance in E-Commerce Websites: Insights and Lessons Learned from Human and LLM Analyses}


\author{Eman Alashwali}
\authornote{Corresponding author and principal investigator.}
\affiliation{%
  \institution{King Abdulaziz University}
  \city{Jeddah}
  \country{Saudi Arabia}}
\email{ealashwali@kau.edu.sa}

\author{Abeer Alhuzali}
\affiliation{%
  \institution{King Abdulaziz University}
  \city{Jeddah}
  \country{Saudi Arabia}}
\email{aalhathle@kau.edu.sa}
\renewcommand{\shortauthors}{}

\begin{abstract}
In 2024, Saudi Arabia’s Personal Data Protection Law (PDPL) came into force. However, little work has been done to assess its implementation. In this paper, we analyzed 100 e-commerce websites operating in Saudi Arabia against the PDPL, examining the presence of a privacy policy and, if present, the policy's declarations of four items pertaining to personal data rights and practices: \begin{inparaenum}[1)] \item personal data retention period, \item the right to request the destruction of personal data, \item the right to request a copy of personal data, and \item a mechanism for filing complaints\end{inparaenum}. Our results show that, despite national awareness and support efforts, a significant fraction of e-commerce websites in our dataset are not fully compliant with the examined PDPL requirements at the privacy policy level: only 31\% of websites in our dataset declared all four examined items in their privacy policies. Even when privacy policies included such declarations, a considerable fraction failed to cover required fine-grained details. Second, the majority of top-ranked e-commerce websites in our dataset (based on search results order) and those hosted on local e-commerce hosting platforms exhibited significantly higher non-compliance rates than mid- to low-ranked websites and those not hosted on local e-commerce platforms. Third, we explored the use of Large Language Models (LLMs) as an automated tool for privacy policy analysis to measure compliance with the PDPL, using the GPT-5 model. We highlight the potential of LLMs and suggest considerations that can inform future benchmarks aimed at improving LLM-based automated analysis for privacy policies. Our results provide a step forward in understanding implementation barriers to data protection laws, especially in non-Western contexts. We provide recommendations for policymakers, regulators, website owners, and developers seeking to improve data protection practices and automate compliance monitoring.
\end{abstract}

\keywords{PDPL, privacy, policy, law, regulation, compliance, LLM, data, retention, destruction, copy, complaint}

\maketitle
\begin{tcolorbox}{This is the author's preprint for a paper accepted at the IEEE Access journal. Please cite and refer to the official version at: https://ieeexplore.ieee.org/document/11675971. DOI: 10.1109/ACCESS.2026.3730247}
\end{tcolorbox}
\renewcommand{\shortauthors}{Alashwali and Alhuzali}

\section{Introduction}
\subsection{Background and Motivation}
Most countries worldwide have recognized the importance of personal data protection and have enacted laws and regulations accordingly. To list a few, the European General Data Protection Regulation (GDPR)~\cite{gdpr25}, the California Privacy Rights Act (CPRA)~\cite{cpra20}, and China's Personal Information Protection Law (PIPL)~\cite{pipl21}. 

More recently, Saudi Arabia introduced its Personal Data Protection Law (PDPL)~\cite{uae25}\footnote{The abbreviation PDPL may be used by other countries to refer to a similar law (e.g., the United Arab Emirates' PDPL~\cite{uae25}). In this paper, we use the term PDPL to refer to the Saudi PDPL~\cite{sdaia23}.}, which was fully enforced in 2024. These legal frameworks commonly aim to safeguard individuals' personal data, regulate how organizations process users' data, and promote accountability and transparency in data handling practices.

To comply with these legal requirements, data controllers (website owners) must be transparent about their data practices. Article 4 of the PDPL regulation explicitly mandates data controllers collecting data directly from data subjects (website users) to inform them of their rights and the mechanisms to exercise these rights~\cite{sdaia23_1}. One of the principal means of ensuring such transparency is through a privacy policy, which serves as a public declaration of how data controllers process personal data (e.g., collect, use, retain, share, and destroy it). The PDPL, in particular, requires controllers to make their privacy policies accessible and to clearly communicate data subjects' rights and relevant procedures. 

Despite this regulatory progress, privacy concerns among users, both in Saudi Arabia and globally, remain prevalent~\cite{mcdonald10,zhang10,alsulaiman14,alsagri15,alashwali24}. Only a few studies have examined the PDPL enforcement and compliance by analyzing privacy policies on Saudi websites~\cite{alhazmi25,mashaabi23}. These studies either narrowly assessed the presence of privacy policies and cookie consent~\cite{alhazmi25}, or evaluated Machine Learning (ML) models to analyze high-level properties~\cite{mashaabi23}. To our knowledge, none of these studies analyzed Saudi e-commerce websites at a fine-grained level of detail, despite their extensive personal data processing, or explored the feasibility of using Large Language Models (LLMs) as supportive tools for privacy policy analysis against the PDPL.

To bridge this gap, we analyzed 100 e-commerce websites operating in Saudi Arabia (i.e., collecting and processing personal data for data subjects residing in Saudi Arabia) to assess their compliance with key PDPL requirements at the privacy policy level. We examined the existence of a privacy policy and, if present, conducted a fine-grained analysis of four policy items, selected for their perceived importance, namely: the policy's declarations of: \begin{inparaenum}[1)] \item personal data retention period, \item the right to request the destruction of personal data, \item the right to request a copy of personal data, and \item a mechanism for filing complaints\end{inparaenum}. Furthermore, we assessed the effectiveness and limitations of LLMs as automated tools for privacy policy analysis compared to human analysis.

\subsection{Research Questions}
Our work aims to answer the following research questions: 
\begin{itemize}
    \item \textbf{RQ1:} How compliant are e-commerce websites operating in Saudi Arabia with the PDPL's privacy policy requirements? 
    \item \textbf{RQ2:} How do websites' ranking and e-commerce hosting platforms relate to compliance with the PDPL's privacy policy requirements in e-commerce websites operating in Saudi Arabia? 
    \item \textbf{RQ3:} How reliable are LLMs when used as an automated analysis tool for assessing privacy policy compliance with the PDPL requirements compared to human analysis?
\end{itemize}

\subsection{Contributions and Key Findings}
This paper makes the following key contributions:
\begin{itemize}
    \item \textbf{We present one of the first in-depth empirical analyses of Saudi websites' privacy policy compliance with the PDPL.} Our analysis reveals that, to this writing, Saudi e-commerce websites remain far from compliant with the PDPL's privacy policy requirements: 9\% of websites failed to provide a privacy policy, 31\% provided privacy policies that failed to declare any of the four examined items, 29\% declared some, but not all of the items, and only 31\% declared all of them. 
    
    \item \textbf{We go beyond coarse-grained privacy policy analysis, providing a fine-grained assessment of mandated details.} Our fine-grained privacy policy analysis shows that even when the examined privacy policy items are declared, a considerable fraction of privacy policies omit required fine-grained details. For example, among the privacy policies that declared the right to request the destruction and a copy of personal data, or a mechanism for filing complaints, more than 80\% lacked a time frame for responding to those requests. 
    
    \item \textbf{We provide previously unknown insights on the relationship between websites' ranking, e-commerce hosting platforms, and compliance.} Our e-commerce websites' analysis shows that top-ranked websites (based on search results order) and those hosted in local e-commerce platforms tend to have significantly higher non-compliance rates compared to mid- and low-ranked websites and those not hosted on local e-commerce platforms, with 60\% of the top-ranked websites and 70\% of the websites hosted on local e-commerce platforms being categorized as non-compliant.
    
    \item \textbf{We assess the use of LLMs as an automated privacy policy analysis tool and compare an LLM's answers to coarse-grained privacy policy questions to those of human analysts.} Our comparative analysis of human and LLM answers to coarse-grained privacy policy questions shows agreement rates of over 80\% on 3 of 4 questions. Our analysis shows that the LLM outperformed human accuracy in some cases, particularly in finding policy items' declarations in overly long privacy policies and in those that lack sectioning or use irrelevant section headings. However, our analysis also suggests considerations for LLM-based privacy policy analysis, including aligning HTML inner and outer privacy policy content and specifying the boundaries of the privacy policy text by website developers. 
\end{itemize} 

Our findings provide actionable recommendations for policymakers, regulators, website owners, and developers seeking to improve transparency, automate compliance monitoring, and enhance data protection practices.

\section{Background and Related Work} \label{sec:related}

\subsection{Data Protection Laws and the Saudi's PDPL} \label{sec:priv_law}
Most countries worldwide have recognized the importance of personal data protection. According to statistics from the United Nations (UN) trade and development~\cite{unatd25}, as of 2025, 79\% of countries worldwide have data protection and privacy legislation. For example, 
the European General Data Protection Regulation (GDPR) (2018)~\cite{gdpr25}; the US's California Consumer Privacy Act (CCPA) (2020)~\cite{ccpa18} and California Privacy Rights Act (CPRA) (2023)~\cite{cpra20}; the China's Personal Information Protection Law (PIPL) (2021)~\cite{pipl21}; and the Singapore's Personal Data Protection Act (PDPA) (2012 amended in 2021)~\cite{pdpa25}, to list a few~\cite{dla25}.

More recently, Saudi Arabia implemented the Personal Data Protection Law (PDPL), issued in 2021, amended in 2023, and took effect in 2023 with a one-year transition period before being fully enforced in 2024~\cite{sdaia23}. This legislation regulates the processing of personal data, such as collecting, storing, modifying, using, disclosing, sharing, and destroying data, by any means, whether manual or automated, regardless of data format, source, or type~\cite{sdaia23,sdaia23_2}. Its jurisdiction extends to all personal data processing activities within Saudi Arabia and to the data of Saudi residents processed abroad~\cite{sdaia24}. The PDPL obliges data controllers to provide a privacy policy (Article 12)~\cite{sdaia23}. Moreover, under Article 4 of the PDPL, data subjects have five fundamental rights, including the right to be informed of the purpose and legal basis for collecting their data, and the right to access, request, correct, and destroy their personal data held by the controller~\cite{sdaia23}. Along with the law document, the Saudi Data and Artificial Intelligence Authority (SDAIA) issued the implementing regulation of the PDPL~\cite{sdaia23_1}, which provides further details, and a privacy policy guideline~\cite{sdaia24}. The law and the implementing regulation are regulatory references, while the privacy policy guideline is not a binding regulatory document~\cite{sdaia24}. 

\subsection{Privacy Policies} \label{sec:priv_policy}
Websites use privacy policies to communicate whether and how they process personal data, e.g., collect, store, share, and destroy it. On the web, privacy policies are commonly presented in natural language text. However, natural-language text-based privacy policies have long been known to be time-consuming and difficult to read and comprehend~\cite{hochhauser01,graber02,pollach07,mcdonald08,reidenberg15,fabian17}. To address these challenges, several solutions have been proposed to improve the presentation of privacy policies. One of the earliest solutions was the Platform for Privacy Preferences (P3P), a standardized computer-readable format for website privacy policies~\cite{cranor25}, which enabled automated privacy policy analysis tools that signal a website's privacy practices against a user's privacy preferences, such as the Privacy Bird browser extension~\cite{cranor06}, and the Privacy Finder search engine~\cite{byers04}. Other solutions proposed different presentation styles for privacy policies that go beyond plain text, such as the expandable grid to visualize P3P privacy policies~\cite{reeder08}, nutrition labels for privacy~\cite{kelley09}, the Poli-see (an interactive tool for visualizing privacy policies)~\cite{guo20}, privacy icons~\cite{holtz11}, and privacy rating visualization~\cite{barth21}, to list a few.

Despite numerous efforts to improve the presentation of privacy policies, none have gained widespread adoption in practice, and natural-language text-based privacy policies remain the norm on today's web. Thus, improving the analysis of natural-language text-based privacy policies has gained considerable attention from researchers.

\subsection{Privacy Policy Analysis} \label{sec:priv_analysis}
Research on privacy policy analysis can be divided into two main themes: manual and automated analysis.

\subsubsection{Manual Analysis}
Multiple studies have examined or employed manual approaches to privacy policy analysis, demonstrating their applicability and usefulness for summarizing data practices and uncovering policy issues at small scales. Wilson et al. found that crowdsourcing privacy policy annotation is a viable means of analyzing privacy policies and yielded high-accuracy annotations~\cite{wilson16}. Shvartzshnaider et al. introduced a privacy policy annotation framework based on the Contextual Integrity framework to reason about privacy in terms of data-flow appropriateness~\cite{shvartzshnaider19,nissenbaum09}. They evaluated their framework with 141 crowdworkers who annotated excerpts from a set of privacy policies. Their approach also achieved high annotation accuracy. Habib et al. manually analyzed the usability of the data deletion and opt-out options on 150 websites in the US using manual content analysis, utilizing a survey completed by the researchers~\cite{habib19}. This approach allowed them to uncover usability issues in privacy policies. Numerous studies adopted manual analysis approaches to analyze privacy policies across sectors and applications, such as~\cite{earp05,winkler16,papacharissi05}, to name a few. 

While helpful, manual approaches for privacy policy analysis are difficult to scale. Thus, research aimed at automating privacy policy analysis has emerged. 

\subsubsection{Automated Analysis}
In what follows, we summarize two significant lines of research in the automated analysis of natural-language text-based privacy policies: ML- and LLM-based approaches.

\par{\textbf{ML-based analysis.}} Numerous studies proposed ML-based solutions for natural-language text-based privacy policies. Zimmeck and Bellovin introduced Privee, an ML classifier augmented by crowdsourcing and implemented as a browser extension~\cite{zimmeck14}. It displays a summary of key data practices and a letter grade for a website's policy. Multiple studies have focused on the automated extraction of specific choices using ML techniques, such as the opt-out choice in privacy policies ~\cite{sathyendra16,kumar20,sathyendra17}. Harkous et al. introduced Polisis, an automated privacy policy analysis framework built on neural network classifiers that enable both structured queries over policy text and free-form queries~\cite{harkous18}. Andow et al. introduced PolicyLint, a tool that identifies contradictory statements about data sharing and collection in a privacy policy using ontology and NLP techniques~\cite{andow19}. 
Building on PolicyLint~\cite{andow19} and AppCensus (a service for Android app traffic analysis)~\cite{census25}, Andow et al. introduced Policheck, a tool to check the consistency of policy statements with their corresponding network traffic~\cite{andow20}. 
Cui et al. introduced POLIGRAPH, a framework that leverages knowledge graph techniques to capture more policy context, identify common patterns across multiple policies, assess term correctness, and detect contradictions with higher accuracy than PolicyLint~\cite{cui23}. 

While ML-based privacy policy analysis techniques enabled analysis at scale, they are costly to develop and run~\cite{rodriguez24_a}. Thus, research leveraging LLMs in privacy policy analysis has gained considerable traction recently.

\par{\textbf{LLM-based analysis.}} With the introduction of LLMs, several researchers examined their use to aid privacy policy analysis at scale. Rodriguez et al. examined the GPT-4 model to analyze the privacy policies of \num{2235} Android apps with respect to data retention declarations~\cite{rodriguez24}. 
Another study by Rodriguez et al. examined different LLM tools (ChatGPT and Llama 2), prompts, parameters, and approaches to identify which achieve the best performance for analyzing privacy policies~\cite{rodriguez24_a}. Sun et al. designed a GPT LLM-agent that serves as an expert system, summarizing key aspects of data practices and answering users' questions about a privacy policy interactively~\cite{sun25}. They conducted a 100-participant survey that suggests that this approach significantly improved user comprehension of privacy policies and reduced task time. Huang et al. tasked GPT-4 Chatbot to annotate segments of \num{2545} corporate privacy policies covering various data practices~\cite{huang24}. They compared the LLM annotation to the human annotation and obtained an overall high accuracy for the examined practices and rights. Freiberger et al. designed PRISMe, an LLM-backed Chrome extension that rates and summarizes privacy risks in a privacy policy and includes an interactive chat~\cite{freiberger25}. While participants found the tool helpful for simplifying privacy policies, they expressed concerns about trusting its ratings. Xie et al. used LLM for analyzing privacy policies at scale (100K websites), achieving higher performance than prior work 
across all tasks~\cite{xie25}. However, the aforementioned studies are in Western contexts. 

Few studies provided insights into LLMs' performance on non-Western contexts. Mori et al compared the comprehension of Japanese privacy policies between an LLM and human participants~\cite{mori25}. The correct answer rates were 85.2\% for LLMs and 63\% for humans, suggesting that LLMs can surpass humans, especially in understanding technical terms. They also evaluated the LLM to identify the data types collected and processed by a privacy policy on 19 Android mobile applications. The best F-1 score they obtained (evaluation per app) was 0.57. Octora et al. introduced CI-PDP, a framework that integrates the Contextual Integrity framework with Indonesia's Personal Data Protection Law (PDP) requirements~\cite{octora25}. They examined LLMs to measure the compliance of three Indonesian privacy policies with 11 pre-defined CI-related attributes. Their results suggest that LLMs performed well on basic parameters but declined on implicit or abstract ones, with Perplexity outperforming ChatGPT across macro-averaged metrics, achieving 0.67 precision, 0.51 recall, and 0.52 F1-score. However, their results should be considered preliminary as they assessed their framework against only three privacy policies. 

The aforementioned results suggest that LLMs show promise for privacy policy analysis at scale. However, our comparison of LLMs and humans identified potential reasons for discrepancies that none of the previous work addressed.

\subsection{Privacy Policy Analysis in the Saudi Arabian Context} \label{sec:priv_saudi}
It has long been known that usable privacy and security research is skewed toward WEIRD (Western, Educated, Industrialized, Rich, and Democratic) samples~\cite{ayako24}. Thus, it is not surprising that there are only a few noteworthy studies on privacy policy analysis for Saudi websites and regulations. Showail et al. manually analyzed 66 mobile apps for Hajj and Umrah (religious activities associated with sacred locations in Saudi Arabia) after the PDPL enforcement~\cite{showail25}. Their study revealed that most apps are not compliant with the PDPL. For example, compliance with users' data protection rights, such as the right to object, access, or delete data, ranged from 18.2\% to 33.3\%. Mashaabi et al. assessed 11 ML algorithms for analyzing privacy policies against compliance with PDPL~\cite{mashaabi23}. The focus of their study was on benchmarking the ML algorithms, and their dataset was collected in 2022, before the PDPL was fully enforced in 2024. It lacks a fine-grained analysis of privacy policies and a sufficient discussion of websites’ data practices, implications, and recommendations to improve the status quo. 

Unlike prior work, we not only examine the presence/absence of privacy practices and rights, but also analyze fine-grained details. We also uncover previously unknown relationships between websites' ranking (based on search results order), e-commerce hosting platforms, and PDPL compliance. Moreover, we explore LLMs as an automated, supportive tool for regulators and data subjects alike to answer basic policy questions. We distill previously unpublished considerations that can inform future benchmarks aimed at improving LLM-based privacy policy analysis.
\section{Methods}\label{sec:methods}
Our method comprises both quantitative and qualitative analyses. Quantitative analysis was conducted using descriptive statistics to count frequencies~\cite{kaur18}. Qualitative analysis was performed by two experienced members of the research team who are native Arabic and fluent English speakers, using template analysis, a style of thematic analysis that combines both inductive and deductive coding~\cite{king24,brooks14}. Privacy policy analysis comprises two stages: a manual analysis, followed by an automated LLM-based analysis. The manual analysis is divided into two phases: coarse-grained, followed by fine-grained, whereas the automated LLM-based analysis is coarse-grained only, as this is an exploratory study of our LLM-based approach. In what follows, we detail our methods.

\subsection{List of E-Commerce Websites} \label{sec:list_method}
\subsubsection{Compiling}
To compile a list of local and global e-commerce websites operating in Saudi Arabia, we used Google web search~\cite{google25} from a device located in Saudi Arabia. To minimize bias, we performed the search using Google Chrome's private mode (Incognito)~\cite{incognito26}, without user sign-in, from a clean-state browser (e.g., cleared browsing data, history, cookies, downloads, cached files, etc.) We aimed for a mixture of local and global websites to mimic an ordinary Saudi-based user's experience, not limited to local websites, and Google search was deemed the most practical approach for building such a list. Most websites operating in Saudi Arabia support both Arabic and English, with one set as the default; however, some websites support only one language. Thus, we considered both languages in our search for e-commerce websites. We utilized Google's advanced search features to find websites operating in Saudi Arabia. We performed two search queries on June 22, 2025. The first was to find websites containing any of the following English keywords: \enquote{purchase}, \enquote{store}, \enquote{shopping}, \enquote{product}, and \enquote{buy}, yielding 180 results. The second was to find websites with any of the following Arabic keywords (the equivalent to the English keywords): \enquote{\ar{شراء}},
\enquote{\ar{متج}},
\enquote{\ar{تسوق}},
\enquote{\ar{منتج}}, 
and \enquote{\ar{اشتر}}
which yielded 227 results. For each query, we extracted all resulting URLs, excluding sponsored results (as they may be less relevant), using the \enquote{URLs Exporter and Importer} Google Chrome browser extension~\cite{ulrs_ext25}. We obtained two lists of URLs: one from the English keywords and the other from the Arabic keywords. We trimmed the URLs to include only the domain name, without the full path.

\subsubsection{Validating} To validate the websites, we manually visited each domain in our lists and checked that it is: functional; an e-commerce website; has a user login; and is operating in Saudi Arabia. The website is considered functional if the homepage successfully loads in the web browser. We considered a website an e-commerce site if it sells goods or services, has a shopping cart, and has payment functionality. To verify that the website has a user login, we searched for a login icon or link. Finally, to check if the website is operating in Saudi Arabia, we looked for indicators of Saudi Arabia (e.g., \enquote{sa}, \enquote{ksa}, or {Saudi Arabia}) in the URL and in the website's content (e.g., homepage's title or language; a language, country, or currency selectors; shipment, contact, or about details). Our dataset included e-commerce websites operating in Saudi Arabia that were either originating from Saudi Arabia (e.g., local businesses) or from outside Saudi Arabia (e.g., global businesses), as both collect and process data for data subjects residing in Saudi Arabia and must adhere to the PDPL~\cite{sdaia24}.

\subsubsection{Cleaning} We removed duplicate entries within the list and between lists. For between-list duplicate entries, we prioritized the Arabic version of the website. We also removed partially duplicate entries that share the base domain but differ in subdomains. For those cases, if both entries have different subdomains, we prioritized the more general/representative subdomain, such as \enquote{www.} and \enquote{en-sa}, and if one of the entries has a subdomain and the other does not, we removed the entry that has a subdomain, as the base domain is more representative of the website. We obtained a total of 160 unique working e-commerce websites: 102 from the Arabic list and 58 from the English list.

\subsubsection{Finalizing} We then merged the two lists and randomly selected 100 websites for analysis. We obtained 63 websites from the Arabic list and 37 from the English list. 

\subsection{Websites' Ranking, E-Commerce Hosting Platforms, and Categories} \label{sec:rank_method}
\subsubsection{Ranking}
We ranked the websites in our dataset according to their order of appearance in the results of the Google search that we performed to compile the lists. We divided the search results in each list (Arabic and English) into three ranks: top, middle, and bottom. See~\Cref{tab:search_rank} in~\autorefappendix{app:method_ext} for the distribution of websites' ranking in our dataset.

\subsubsection{E-Commerce Hosting Platforms} 
For each website, we checked whether it was hosted on an e-commerce platform and, if so, identified the platform provider. To this end, we used the~\enquote{Wappalyzer-Technology profiler} Google Chrome browser extension~\cite{webalyzer25,webalyzer25_2}, which maintains fingerprints for web technologies that the extension matches against, such as HTML, response headers, cookies, etc., and is updated every few weeks\footnote{In a private email communication with the tool's developer, Elbert Alias~\cite{alias26}.}. We visited each website, ensured that the homepage loaded, recorded the hosting platform reported by the extension, if any, and then exported the identified technologies. We performed this process twice to confirm the results. We then divided the e-commerce hosting platforms into three categories: local, non-local, and no-hosting. The local category comprises e-commerce hosting platforms based in Saudi Arabia (namely, Zid~\cite{zid25} and Salla~\cite{salla25}), the non-local category comprises those based outside Saudi Arabia, and the no-hosting category comprises websites that did not indicate use of an e-commerce hosting platform. To determine the country of origin of the e-commerce hosting platform, we conducted a manual Google search and examined the hosting provider's websites and information pages, such as Wikipedia. See~\Cref{tab:hosting_srvs} in~\autorefappendix{app:method_ext} for a breakdown of the identified platforms along with their country of origin.

\subsubsection{Categorizing}
To gain insight into the types of websites included in our dataset, two members of the research team manually categorized the websites at two levels: business model and product type. Each website was visited by the two researchers in the same setting, and for each level of the coding scheme, they assigned code(s) to the website acceptable to both researchers. See~\Cref{tab:categories} in~\autorefappendix{app:method_ext} for the distribution of websites' categories in our dataset. 


\subsubsection{Relationship Between Websites' Ranking, E-Commerce Hosting Platforms, and Compliance}
To identify whether there are relationships between websites' ranking and compliance, e-commerce hosting platforms and compliance, and websites' ranking and hosting platforms, we used the Chi-squared test (denoted as $\chi^2$)~\cite{mchugh13}, given that no more than 20\% of the expected frequencies are less than 5 and none of the cells is less than one, otherwise we used Fisher's exact test~\cite{kim17}. We used an alpha level of .05 for these statistical tests. 
If the p-value for the Chi-squared or Fisher's test was significant (i.e., p $\leq$ 0.05), we conducted a post-hoc analysis using the adjusted residuals to identify the source of significance in our 3 $\mytimes$ 3 contingency tables~\cite{sharpe15,peter26}. We performed corrections for multiple tests using the Bonferroni method. The critical value for the adjusted residuals in a 3 $\mytimes$ 3 table is 2.77. That is, adjusted residuals greater than 2.77 or less than $-2.77$ are identified as those contributing to the significance. All tests were computed using the R statistical tool~\cite{rproject26}. We computed the Chi-squared effect size using Cramer's V (denoted as $\alpha_c$).

\subsection{Manual Analysis of Privacy Policies} \label{sec:manual_analysis}
The manual qualitative analysis is divided into two phases: coarse-grained, followed by fine-grained analysis. We used a survey as a template for analysis. 

\subsubsection{Survey} \label{sec:survey_method}
To guide the manual analysis, we designed a Google Forms survey for the researchers to complete for each website. The survey begins with a checklist to ensure that the researchers are browsing a fresh version of the website and are using Google Chrome's Incognito mode. The first section of the survey asks about basic details of the website's homepage and privacy policy, including the language of the default homepage and privacy policy, whether a working privacy policy is available, and the label of the privacy policy link. This section also asks the researcher to retrieve the content of the website's homepage and privacy policy. The second section captures the high-level content presentation details of the privacy policy (e.g., graphical, textual, or other). The third section captures details of the four policy items we examine. For each item we examine, the survey asks whether the privacy policy informs data subjects about it and asks the researcher to extract the policy section heading(s) and paragraph(s) in which the information is provided. See~\autorefappendix{app:survey} for the full survey.

\begin{table}[t!]
\small
\centering
\begin{threeparttable}
\caption{The policy items we analyzed along with their PDPL requirements we checked~\cite{sdaia23,sdaia23_1,sdaia24}\tnote{a}. In the left column, items followed by ** are explicit rights in the law document~\cite{sdaia23}. Supporting clauses for each item are provided in~\Cref{tab:rights_ext} in~\autorefappendix{app:method_ext}.}
\label{tab:rights}
\begin{tabular}{p{1.7cm}p{6cm}}
\toprule
Item & Requirements \\
\hline
 & The website should ... \\
\hline 
Privacy policy & 
\begin{inparaenum}[1)]
\item Provide a privacy policy and make it available to data subjects before data collection. 
\item The privacy policy should be written in understandable language.
\item The privacy policy should contain other links related to the privacy policy, such as the cookie policy and terms and conditions. 
\end{inparaenum} \\
\hline
 & The privacy policy should ... \\
\hline 
Retention & 
\begin{inparaenum}[1)]
\item Inform data subjects about the retention period or the criteria used to determine that period. \item Provide retention periods for each personal data category. 
\item Provide the method for destroying the data after it is no longer needed.
\end{inparaenum}  \\
\hline

Destruction** & 
\begin{inparaenum}[1)]
\item Inform data subjects of their right to request the destruction of their personal data. 
\item How to exercise this right. 
\item The communication channels for submitting requests. 
\item The time frame for response.
\end{inparaenum}  \\
\hline

Copy** & 
\begin{inparaenum}[1)]
\item Inform data subjects of their right to request a copy of their personal data in a readable and clear format.
\item How to exercise this right. 
\item The communication channels for submitting requests. 
\item The time frame for response.
\end{inparaenum}  \\
\hline 

Complaint & 
\begin{inparaenum}[1)]
\item Provide a mechanism for data subjects to file complaints and objections related to their data rights and processing. 
\item The name of the department or division for handling the complaints. 
\item Its contact details. 
\item The time frame for response. 
\item Name the competent authority in case of unsatisfactory outcome.
\end{inparaenum}  \\
\bottomrule
\end{tabular}
\begin{tablenotes}
\item[a] In August, 2025, we retrieved the latest version of the implementing regulation and found that it differs from the previous version we retrieved in April 2025 and started with. Neither of the two regulation documents has a version or date stamp. We revised our work in accordance with the latest version~\cite{sdaia23_1}. The amendments made to the initial document did not affect our analysis or findings.
\end{tablenotes}
\end{threeparttable}
\end{table}

\subsubsection{Coarse-Grained Manual Analysis} 
Utilizing the survey described in the previous section, each website was visited and qualitatively analyzed by two experienced researchers who submitted a survey response for each website. For each website, the researchers visited the homepage and looked for the \enquote{privacy policy} or similar links containing the word \enquote{privacy}. If not found, they checked other pages where the privacy policy may be located, such as \enquote{terms of use} and login pages. They then answered the survey questions about the homepage and privacy policy (see~\autorefappendix{app:survey}). In addition, for both versions of the website (Arabic and English, if available), one researcher stored the homepage and privacy policy HTML pages using the \enquote{Save Page WE} Google Chrome browser extension~\cite{we23}, or otherwise the browser's \enquote{Save as...} functionality, if the extension failed to save the pages in a few cases.

To ensure a consistent coding process between the two researchers, they first jointly analyzed 9 privacy policies at the beginning of the analysis. Then, they independently analyzed the remaining 91 websites in small batches ($\approx$10 websites per day) such that each website was double-coded. After coding each batch, both researchers met and compared their answers across the websites. In cases of disagreement, they discussed and revisited the website privacy policy as needed until they reached an answer agreeable to both of them. Code definitions were updated throughout the process. The coarse-grained manual analysis was conducted from June 29 to July 10, 2025. The manual analysis concluded when both researchers reached complete agreement on all codes. Nevertheless, to give statistical insights about the level of agreement between the two coders when they coded independently, we retrospectively computed the Inter Reliability Rate (IRR) using Cohen's Kappa ($\kappa$) statistical test for the 91 websites that were coded by both researchers independently\footnote{IRR is not applicable for the first 9 privacy policies as they were coded jointly by both researchers, which entails complete agreement.} before they met to resolve disagreements and update the codebook to reach consensus in all assigned codes. We computed Cohen's kappa for each of the four privacy policy coarse-grained questions and for all four questions combined. Namely, regarding: \begin{inparaenum} \item personal data retention period, \item the right to request the destruction of personal data, \item the right to request a copy of personal data, and \item a mechanism for filing complaints.\end{inparaenum} The overall (\textit{N = 91}) Cohen's Kappa ($\kappa$) scores we obtained per question are (Q1: 0.71; Q2: 0.72; Q3: 0.67; Q4: 0.73) and for all questions combined (Q1 -- Q4: 0.71), which indicates a \enquote{substantial} agreement (0.61 -- 0.80), according to the widely used Landis and Koch's scale~\cite{landis77}. The IRR increased as the coders met and discussed disagreements and code definitions.~\Cref{tab:llm_metric} in~\autorefappendix{app:method_ext} shows the IRR for the first 20\% of the dataset, the remaining 80\%, and the full dataset (reported above).


\subsubsection{Fine-Grained Manual Analysis} 
After completing the coarse-grained coding and having survey responses agreeable to both researchers for all websites in our dataset, the two researchers conducted a fine-grained qualitative analysis using template analysis~\cite{king24,brooks14}. In this phase, the first researcher served as a primary coder, and the second as a reviewer. A single experienced coder is deemed acceptable in template analysis and HCI research~\cite{king24,mcdonald19}, and we supplemented that with a thorough review by a second coder to ensure the reliability of the codes made by the primary coder~\cite{mcdonald19}. Starting from the coarse-grained analysis of survey responses, for each website, the first researcher read through the privacy policy text segments captured in the survey and created an initial codebook (the template) that captured fine-grained details based on the PDPL for each examined policy item and coded the entire dataset. After the first researcher finished coding the entire dataset, the second researcher reviewed the coding. Both researchers discussed and resolved disagreements and updated the codebook as needed during the analysis. Both researchers reached a consensus on the codes.~\Cref{tab:homepage_pres}--~\Cref{tab:complaint} in~\autorefappendix{app:codebook} provide the fine-grained analysis codebook.

\subsection{Automated Analysis of Privacy Policies} \label{sec:auto_analysis}
The automated analysis was conducted only at a coarse-grained level, using a privacy policy text extractor and an LLM agent. 

\subsubsection{Text Extractor}
To provide the LLM agent with privacy policy texts, we extracted only the text (excluding \texttt{script}, \texttt{style}, and \texttt{noscript} tags and their contents) from the locally saved HTML pages for the privacy policies using the \texttt{BeautifulSoup} Python package~\cite{pypl25}. We chose to rely on textual content only because raw HTML files are much larger and contain unnecessary elements such as headers, images, and scripts, which may exceed the LLM's input token window and incur a high cost. 

To count the number of words in a privacy policy, we created a manually cleaned version of each privacy policy text file, which includes only text within the privacy policy boundaries (without the page's header and footer text). Then, we used Python's \texttt{len} function after splitting the text into a list of words using the \texttt{split} function.

\subsubsection{LLM Agent}\label{sec:agent_method}
We developed an LLM agent in Python~\cite{python25} backed by OpenAI's GPT-5 model~\cite{gpt5}. GPT-5 was selected as it represents the most recent release at the time of the analysis, offering enhanced reasoning capabilities, an extended context window of up to 400K tokens, and a knowledge cutoff date of September 30, 2024~\cite{gpt5_a}. In models prior to GPT-5, prior work recommended setting the model's temperature to 0 as this increased its determinism~\cite{rodriguez24_a}. However, the then-new GPT-5, which we used, does not support adjusting the temperature parameter~\cite{robin25}. For our exploratory LLM experiment in this study, we adopted a zero-shot, role-defined prompt (as a privacy compliance expert).

For each privacy policy, the agent constructs a prompt (depicted in~\cref{fig:prompt}) to query the GPT-5 model. The prompt starts with the following instruction: 
\textit{\enquote{You are a privacy compliance expert. Use ONLY the provided policy text below to answer the question. Do not rely on outside knowledge. If the answer is not found, state that clearly}}, followed by the privacy policy's text read from a text file, and a question to be answered by the LLM drawn from a CSV file containing the four privacy policy questions. The prompt is automatically generated for each question in the CSV, ordered from 1 to 4. The questions are coarse-grained (i.e., whether a policy item was declared or not without fine-grained details) multiple-choice questions that the researchers answered in the analysis survey~(\cref{sec:survey_method}), offering three options: \enquote{Yes}, \enquote{No}, and \enquote{Other: please specify}, and focused on whether the policy informs data subjects about a policy item. The four questions are provided in~\autorefappendix{app:llm-questions}. The English prompt was used for both Arabic and English privacy policies.

After executing the prompt, the agent recorded the response for each privacy policy in a CSV file. Each response includes the LLM's answer, reasoning, the policy heading in which the LLM identified the answer, and the LLM's confidence score, ranging from 0 to 1. After collecting all of the LLM's response sheets, we combined the responses into a single sheet. The automated analysis using the LLM agent was conducted on August 13, 2025.

\begin{figure}
\centering
\begingroup
\setlength{\fboxsep}{10pt}   
\setlength{\fboxrule}{0.2mm}  
\fbox{%
\begin{minipage}{\dimexpr\columnwidth-2\fboxsep-2\fboxrule\relax}
\footnotesize
You are a privacy compliance expert. Use ONLY the provided policy text below to answer the question. Do not rely on outside knowledge. If the answer is not found, state that clearly.
\vspace{\baselineskip} 
\par Privacy Policy:
\par """[{privacy policy text}]"""
\vspace{\baselineskip} 
\par Question:
\par [question]
\vspace{\baselineskip} 
\par Options:
\par [option 1]
\par [option 2]
\par [option 3]
\vspace{\baselineskip} 
\par Respond in this format:
\par Answer: <exact option(s)>  
\par Reasoning: <why you selected that answer>  
\par Heading: <policy section>  
\par Confidence: <0.0–1.0>  
\end{minipage}%
}
\endgroup
\caption{The prompt template used by the LLM agent in the automated coarse-grained privacy policy analysis. The LLM retrieves the privacy policy text from a text file, and the question to be answered and its options are retrieved from a CSV file.}

\label{fig:prompt}
\end{figure}
\subsection{\textbf{Comparing Human and LLM Answers}} \label{sec:llm_analysis_method}
After the LLM-agent run completed, two experienced researchers conducted a comparative qualitative analysis of human and LLM privacy policy answers using template analysis~\cite{king24,brooks14}. The first researcher acted as a primary coder while the second acted as a reviewer. A single experienced coder is deemed acceptable in template analysis and HCI research~\cite{king24,mcdonald19}, and we supplemented that with a thorough review by a second coder to ensure the reliability of the codes made by the primary coder~\cite{mcdonald19}. The first researcher compared the researchers' answers for the coarse-grained policy questions with those of the LLM and categorized them as either \enquote{same} or \enquote{different}. For all answers categorized as different, the first researcher reviewed the LLM's and our reasoning and coded the reason for the discrepancy. After the first researcher finished coding the whole dataset, the second researcher reviewed the codes and discussed and resolved any disagreements, if any. The researchers updated the codebook as needed during the analysis. Both researchers reached a consensus on all codes.~\Cref{tab:themes} in~\autorefappendix{app:codebook} provides the comparative analysis codebook.

\subsection{\textbf{LLM Metrics}} \label{sec:llm_metrics_method}
As a first step in evaluating LLM-based privacy policy analysis for a dataset consists of Arabic and English privacy policies, we calculate the accuracy, precision, recall, F1 score, and Cohen's kappa ($\kappa$) of the LLM answers to the coarse-grained privacy policy questions compared to human annotated ground truth (our answers after two analysts reached complete agreement and corrected human errors revealed after we ran the LLM and manually inspected discrepancies~(\cref{sec:human_vs_llm}). The accuracy score reflects agreement; the Macro F1 score is the harmonic mean of precision and recall scores, reflecting the overall balance between correctly answering and avoiding incorrect answers; Cohen's kappa ($\kappa$) score accounts for agreement that occurs by chance. We computed them using the \texttt{sklearn.metrics} Python's module.

\subsection{\textbf{Compliance Definition}}
Our study is at the privacy policy level, measuring partial PDPL requirements from the text. Namely, the existence of a privacy policy and, if present, the privacy policy's declarations of: \begin{inparaenum}[1)] \item personal data retention period (retention), \item the right to request the destruction of personal data (destruction), \item the right to request a copy of personal data (copy), and \item a mechanism for filing complaints (complaint)\end{inparaenum}. For short, we refer to the examined privacy policy items, consequently, as retention, destruction, copy, and complaint. The PDPL requirements we checked for the examined items are listed in~\Cref{tab:rights}.

We categorized websites' compliance into three categories: fully compliant, partially compliant, and non-compliant. We examined the presence of a privacy policy and whether it informs data subjects about retention, destruction, copy, and complaint at a coarse-grained level. That is, whether the policy declares them or not, without examining the fine-grained details. We considered an item to be declared if the manual analysis answer to the item's question is \enquote{yes} or \enquote{other} (the latter typically used for edge cases).

For the items we examined at the privacy policy level, we define fully compliant websites as those that provide a privacy policy that informs data subjects about the four examined items. Partially compliant websites are those that offer a privacy policy that informs about some, but not all, of the four items. Non-compliant websites are those that either do not provide a privacy policy or provide one but do not declare any of the four items we examine. 

In our context, the data controller is the website owner, and the data subject is the website user.

\subsection{Ethical Considerations} \label{sec:ethics}
Our study does not involve interaction with human subjects. Therefore, we did not need Institutional Review Board (IRB) approval. We needed to visit websites and extract their privacy policy text for analysis. Our visits to the websites were nearly equivalent to those of an ordinary consumer using an ordinary web browser, and they did not exhaust the websites' resources. We saved the HTML privacy policy pages manually using legitimate means (a browser extension and the browser's \enquote{Save as...} functionality) and extracted and processed the text offline. Downloading privacy policies and extracting their texts for analysis is a common practice in prior research~\cite{wilson16,habib19,srinath21,srinath24,sun25}, and the only known practical way to conduct this type of research. The privacy policy texts we used are publicly available informational documents that do not require authentication to access and contain no private data. We configured the GPT API not to use our data to train models, to maintain the privacy of the data used with the API. We refrain from sharing the HTML or plain text of the privacy policies publicly to avoid infringing copyrights, if any. Since our study does not involve security vulnerabilities, disclosure to any party was not required. 

\subsection{Limitations} \label{sec:limitations}
First, some websites in our dataset may change their privacy policies or become unavailable over time. Moreover, the reported relationships reflect the time of data collection and may not hold in the future. However, this is a common limitation of web measurement studies and does not diminish the value of the findings for understanding the present and future. The compliance assessment and the relationships we identified are meant to surface current issues to affected parties, so that they can act upon them. Second, we acknowledge that LLMs may have a margin of non-determinism (may change their answers for the same questions between different runs)~\cite{rodriguez24_a}. Moreover, we only examine one LLM (GPT-5) on a single run. This is deemed acceptable for our exploratory study. Third, using Google's search engine to build the websites' sample may have introduced biases, and the Google search engine's underlying algorithms that rank websites in the results are not publicly known. Nevertheless, using Google search to build a list of e-commerce websites was deemed the most practical and applicable approach, as we are not aware of a reliable public directory that lists global and local e-commerce websites operating in Saudi Arabia, let alone one that provides a means of ranking them. We eliminated bias by using a clean-state browser in private mode (Incognito) and using general keywords related to e-commerce and not related to business models, products, or brands as described in. Fourth, our use of the Wappalyzer Chrome browser extension~\cite{webalyzer25,webalyzer25_2} to identify the e-commerce hosting platform that a website uses has known limitations. It relies on fingerprints to identify the technologies used by websites. However, websites can hide or manipulate the technologies they use to evade profiling~\cite{verneaut26}. Moreover, the extension's list of e-commerce platforms may not include every existing platform. However, similar profiling tools are very likely to share these limitations. As of this writing, the Wappalyzer extension has a rating of 4.6/5 (2K ratings) and 3M users~\cite{webalyzer25,webalyzer25_2}, and has been used in prior studies such as~\cite{paniagua25}. Given the exploratory nature of our study, Wappalyzer was deemed sufficient, with an acknowledgment of a potential margin of error in the e-commerce hosting platforms. Fifth, our analysis is limited to the existence of a privacy policy and four privacy policy items on 100 websites. The small dataset may affect the generalizability of the results and the statistical power of the statistical tests we performed. However, this was also deemed sufficient for our exploratory study. Our work includes manual analysis, and manually analyzing a larger set of websites at a fine-grained large-scale level is time- and labor-prohibitive for the project's scope and budget. Manual analysis provides depth at the cost of breadth. This allowed us to identify potential reasons for discrepancies between human and LLM-based analysis that none of the previous work we are aware of has identified~(\cref{{sec:reasons}}), which can improve future benchmarks. Sixth, our compliance measurement assessed data controllers' privacy policies, not their actual behavior. Thus, the term \enquote{compliance} should be interpreted in this context. That is, at the privacy policy level, for its existence and the declaration of the four examined policy items. Seventh, similar to any qualitative analysis, subjectivity is a threat. To mitigate this, two experienced cybersecurity and privacy researchers were involved in all stages of the qualitative analysis. Furthermore, we employed an LLM and compared our answers with those produced by the LLM. Potential sources of discrepancies were analyzed and reported as part of our contributions to further our understanding of LLM-based automated privacy policy analysis.

\section{Results}\label{sec:results}
In this section, we summarize our results. Our dataset comprises websites and their privacy policies, both in English and Arabic. However, in this paper, we summarize both in English. The frequencies are presented as \#/\#, where the top number indicates the number of occurrences, and the bottom indicates the total number of instances considered, followed by the percentage (\#\%).

\subsection{Overview of Analyzed Websites}\label{sec:websites}
Our list of 100 e-commerce websites includes domains of varying ranks. Overall, 35/100 (35\%) websites were in the top, 36/100 (36\%) in the middle, and 29/100 (29\%) in the bottom third of the Google search results. 

Regarding e-commerce hosting platforms, 56/100 (56\%) websites were hosted on e-commerce platforms: 27/100 (27\%) on local platforms, 29/100 (29\%) on non-local platforms\footnote{One website (for a large telecommunication company) was identified as hosted on Amazon WebStore (non-local). However, this platform is deprecated~\cite{pixel26,failory26}.}, and 44/100 (44\%) were not hosted on any platform. 

In terms of websites' categories, according to a two-level classification: business model and product type(s): the business model and product type(s). At the business model level, the most common category was \enquote{retail} (75\%), followed by \enquote{service} (22\%), and \enquote{marketplace} accounted for only 3\% of the websites. At the product-type level, the top four categories of our websites were (each category is surrounded by double quotes, and some categories combine multiple related products. Some websites fell into multiple categories): \enquote{beauty} (18\%), \enquote{appliances/electronics/gadgets} (17\%), \enquote{fashion/apparel} (16\%), \enquote{sports/outdoor} (9\%). See~\Cref{tab:search_rank} --~\Cref{tab:categories} in~\autorefappendix{app:method_ext} for the full breakdown of websites' rankings, distribution across identified e-commerce hosting platforms, and categories.

Regarding the display language(s) of the websites, we examined support for Arabic and English. Among the websites included in our analysis, 99/100 (99\%) offered an Arabic version of the homepage, 78/100 (78\%) provided an English homepage, and 77/100 (77\%) offered both Arabic and English versions of the homepage, while 22/100 (22\%) homepages were exclusively provided in Arabic. Only one homepage was offered solely in English. Browsing from a device located in Saudi Arabia, 71/100 (71\%) homepages defaulted to Arabic, while 29/100 (29\%) defaulted to English. Finally, regarding privacy policies, among the websites we analyzed, 91/100 (91\%) provided one. In the following sections, we present the results of our analysis of privacy policies.

\subsection{Presentation of Privacy Policies}\label{sec:pres}
\subsubsection{Language}
Of the websites that provided a privacy policy, 65/91 (71\%) offered both Arabic and English versions of the privacy policy, 24/91 (26\%) offered only an Arabic policy, and 2/91 (2\%) policies were exclusively in English. In terms of the default language of the homepage for the websites that provided a privacy policy, 63/91 (69\%) defaulted to Arabic, while 28/91 (31\%) defaulted to English. In terms of the language of the privacy policy content, 58/91 (64\%) defaulted to Arabic, while 33/91 (36\%) defaulted to English. In 86/91 (95\%) of the policies, the privacy policy defaulted to the same language as the homepage from which it was accessed. The five inconsistent cases were for homepages that defaulted to Arabic while their privacy policies defaulted to English. Further details about the analyzed websites are provided in~\Cref{tab:summary_ext} in~\autorefappendix{app:results_ext}.

\subsubsection{Link}
In 88/91 (97\%) of the privacy policies, the link to the privacy policy was a direct, one-click link on the homepage, while 3/91 (3\%) provided it as an indirect link within other pages, requiring more than one click from the homepage, namely within the \enquote{help}, \enquote{user policy}, and \enquote{login} pages. Across the 91 analyzed privacy policies, the privacy policy link was labeled in the same language as the homepage, with 59/91 (65\%) labeled as \enquote{privacy policy}, while 32/91 (35\%) used alternative labels such as \enquote{privacy \& security policy}, \enquote{privacy}, and \enquote{privacy \& cookie policies}. In 4/91 (4\%) websites, the direct link to the privacy policy did not contain the word \enquote{privacy} at all, such as \enquote{terms of use}, \enquote{policies}, and \enquote{terms \& conditions}.

\subsubsection{Content Presentation}
In terms of privacy policy content presentation, all 91 privacy policies were presented textually, none included noteworthy graphical content, and only a few included one or more elements to organize the textual content, such as accordion menus. For example, only 10/91 (11\%) included a table of contents, 28/91 (31\%) included additional links related to privacy policy, such as cookie policy, and 4/91 (4\%) included accordion menus, while 58/91 (64\%) privacy policies did not include any of the above or any other noteworthy elements to organize the text.

\subsubsection{Text Relevance}
With respect to privacy policy text relevance to privacy, of the privacy policies reviewed, 19/91 (21\%) were combined with unrelated policies (excluding cases where they were paired with related policies like cookie or security policy), such as \enquote{privacy and terms of use policy}, \enquote{terms \& conditions} and \enquote{rules, privacy policy and refunds}. Of the 19 combined policies, four were presented only as an item. Another 4/91 (4\%) privacy policies contained content that was entirely irrelevant to privacy, such as \enquote{return policy}, \enquote{sales terms and conditions}, and a placeholder: \enquote{Data privacy is being updated}, although the privacy policy link label included the word \enquote{privacy}. The remaining 68/91 (75\%) were presented as standalone privacy policies with relevant content.

\subsubsection{Length} \label{sec:length}
In terms of privacy policy text length (excluding text outside the policy text boundaries such as the page's header and footer text), for all the 91 analyzed privacy policies, the word count ranged from 19 to \num{11761} words, with an average word count of $\approx\num{1943}$ words per policy, and a standard deviation of $\approx\num{2197}$, which is greater than the average. This indicates that the policies' word counts are dispersed and vary substantially in length.

\subsection{RQ1: Practices and Rights in Privacy Policies}\label{sec:rights}
In this section, we summarize the results of privacy policy analysis with respect to the four items we examine (retention, destruction, copy, and complaint). The results presented in this section are after correcting human errors discovered after running the LLM and conducting the comparative analysis for discrepancies between human and LLM answers~(\cref{sec:human_vs_llm}).

\subsubsection{Personal Data Retention Period}\label{sec:retention}
Out of 91 policies, 41/91 (45\%) policies declared the personal data retention period, and 1/91 (1\%) mentioned retention only for a specific category of data (namely, \enquote{comments and their metadata}), while the remaining 49/91 (54\%) did not specify any.

We further categorized the data retention declarations. Out of the policies that declared retention period, 42/42 (100\%) provided the criteria to determine the retention period (e.g., \enquote{as long as necessary for ...}, \enquote{as long as permitted by law}, \enquote{as long as it is required}, etc.), while only 8/42 (19\%) provided specific periods (days, months, years) in addition to retention criteria, e.g., \enquote{for a period of ten years and as long as necessary}. 

We also analyzed the scope of these declarations. Of the 42 policies that declared a retention period or the criteria to determine it, 13/42 (31\%) policies provided retention periods for multiple data categories (e.g., account data, personalization data, and customer support data), while 29/42 (69\%) policies did not specify data categories and referred to personal data in general or a single type of data such as account data and interaction data. 

Finally, we examined whether the policies specified methods for data destruction upon expiration of the retention period. We found that only 10/42 (24\%) stated that the data will be destroyed \enquote{securely}, with 6/10 adding \enquote{promptly}, and 2/10 adding that the data may be \enquote{anonymized} as an alternative to secure destruction after the retention period has expired. The remaining 32/42 (76\%) policies did not specify any destruction method upon retention period expiration.

\subsubsection{Request the Destruction of Personal Data}\label{sec:destruction}
We found that 46/91 (51\%) policies informed data subjects about the right to request the destruction of their personal data, 4/91 (4\%) specified the right to destroy the account or profile data only, and 1/91 (1\%) associated it with the condition \enquote{if you do not agree with the changes}, while 40/91 (44\%) policies did not declare this right.  

We then analyzed the methods for exercising this right in the privacy policies. Out of the 51 policies that granted the destruction right, 36/51 (71\%) described one or more methods for exercising this right, while the remaining 15/51 (29\%) did not provide any. Of the policies that defined methods, 26/36 (72\%) provided method details within the relevant text, with sending an email (16), logging into the user account (11), and submitting a web form (8) as the top three methods, consequently and non-exclusively. Alternatively, 10/36 (28\%) policies exclusively vaguely referred the user to contact the data controller. The contact details were in the privacy policy itself, e.g., in the contact details section of the policy in 7/10 cases, and/or were placed outside the policy boundaries, such as in the page's footer (4/10).  

Finally, we examined the time frame for responding to requests for data destruction in policies that grant such rights. The majority, 42/51 (82\%), did not give a time frame. Only 3/51 (6\%) specified a fixed period (e.g., within 30 days), while 6/51 (12\%) used statements such as \enquote{within a reasonable time}. 

\subsubsection{Request a Copy of Personal Data}\label{sec:copy}
Of the 91 policies we analyzed, 31/91 (34\%) explicitly allowed users to request a copy of their personal data, while 19/91 (21\%) acknowledged the \enquote{right of access to personal data} (Art.5)~\cite{sdaia23_1}. Unlike the GDPR, in the PDPL, the \enquote{right of access} (Art.5) is different from the \enquote{right to request access to personal data} (Art.6)~\cite{sdaia23_1}, where only the latter states that it grants data subjects the right to \enquote{request a copy of their Personal Data in a readable and clear format}~\cite{sdaia23_1}. The remaining 41/91 (45\%) policies did not mention the right to obtain a copy of or access to personal data. 

We then examined the methods that privacy policies provided to enable data subjects to exercise the right to obtain a copy of or access to their personal data. Out of 50 policies, 35/50 (70\%) declared one or more methods for exercising the right. The methods varied, with 32/50 (64\%) policies describing the methods directly in the relevant text. The most commonly mentioned approaches (often in combination rather than exclusively) included logging into the user account (17), sending an email (13), and submitting a web form (7). Alternatively, 3/50 (6\%) exclusively vaguely referred the reader to contact the controller directly. The controller's contact details were embedded within the privacy policy text itself (3/3) and/or placed elsewhere, such as in the policy page's footer (1/3). 

We also examined the timeline for processing requests for a copy of personal data. Only 2/50 (4\%) specified a fixed period (e.g., within 30 days), while 5/50 (10\%) used general time statements such as \enquote{within a reasonable period} or \enquote{promptly}. The majority of policies, 43/50 (86\%), provided no time frame.  

\subsubsection{Mechanism to Filing Complaints}\label{sec:complaint}
We investigated the mechanisms for filing complaints in the analyzed privacy policies. Overall, 30/91 (33\%) policies outlined a mechanism for data subjects to file complaints, while 6/91 (7\%) mentioned the \enquote{right to object} (typically found in the rights section) without a dedicated complaints section. The remaining 55/91 (60\%) policies did not declare any complaint mechanism.

The presentation of complaint information was inconsistent. Of the 36 policies with a mechanism, 17/36 (47\%) included a dedicated section for complaints, whereas 19/36 (53\%) embedded this information in other parts of the policies without a distinct heading.  

Regarding the declaration of the responsible department, division, or officer name for handling the complaints, only 12/36 (33\%) of the policies that included a complaint mechanism specified the name of the department and/or officer responsible for handling data subjects' complaints or objections related to personal data, whereas 24/36 (67\%) omitted this detail.

The channels used to file complaints regarding personal data varied. The e-mail was the most frequently mentioned channel, appearing in 22/36 (61\%) policies, followed by the web form in 6/36 (17\%) and postal mail in 6/36 (17\%), non-exclusively. While directing users to contact the competent authority (the regulator) should be used only when an unsatisfactory outcome is reached with the controller, we find it mentioned as the first channel for filing a complaint in 5/36 (14\%) policies. 
 
Regarding the response time for processing complaint requests, the vast majority of policies, 32/36 (89\%), did not indicate a response time. Only 4/36 (11\%), with 3 of the 4 using general statements (e.g.,\enquote{in time} and \enquote{as soon as possible}) while only one policy specified a concrete time frame (\enquote{within 30 days}). 

Finally, regarding the escalation procedure if the initial complaint remains unresolved, 13/36 (36\%) listed the option to contact the data authority, with 7/13 referring to unnamed data authority (e.g.,\enquote{data protection authority}), while 6/13 named the authority (e.g., SDAIA), with 3/13 (either specific or general) indicated a non-Saudi authority indicated either by name, e.g., the UK's \enquote{Information Commissioner's Office (\enquote{ICO})}~\cite{ico26}, or through the provided contact details for the authority. The remaining 23/36 (64\%) policies did not specify an appeal procedure in the event that a complaint is unresolved by the data controller.

\subsection{RQ2: Websites' Ranking, E-Commerce Hosting Platforms, and Compliance}\label{sec:relations}
\subsubsection{Websites' Compliance.}
Overall, at the privacy policy level, of the 100 analyzed websites, only 31/100 (31\%) were fully compliant (declared all four examined items), 29/100 (29\%) were partially compliant (declared one or more, but not all the examined items), while 40 (40\%) websites were non-compliant (provided no privacy policy or one that failed to declare all examined items).

\subsubsection{Websites' Ranking and Compliance.} We examined whether the ranking of the analyzed websites in our dataset (according to their appearance in Google search results) is related to their compliance with the PDPL. Surprisingly, top-ranked websites in our dataset exhibited the lowest compliance rate: 21/35 (60\%) were non-compliant, compared to 8/36 (22\%) of mid-ranked websites and 11/29 (38\%) of low-ranked websites. The next two sections may provide further insights into this phenomenon. The relationship between websites' ranking and compliance is statistically significant ($\chi^2$ = 11.7, df = 4, p-value = 0.019, $\alpha_c$ = 0.24). A post hoc analysis of the adjusted residuals with a Bonferroni correction suggests that the observed count of top-ranked websites that were classified as non-compliant is significantly higher than expected (adjusted residual $= 3.00 > 2.77$).~\Cref{tab:rank_compliance} in~\autorefappendix{app:results_ext} presents a breakdown of websites' ranking in relation to compliance. 

\subsubsection{E-Commerce Hosting Platforms and Websites' Compliance.} We also examined whether the e-commerce hosting platforms of the analyzed websites are related to their websites' compliance. Overall, we find that websites hosted on local e-commerce hosting platforms had the highest non-compliance rate, 19/27 (70\%). In contrast, websites hosted on non-local and no-hosting platforms had significantly lower non-compliance rates, 6/29 (21\%) and 15/44 (34\%), respectively. Surprisingly, none of the websites hosted on local e-commerce platforms were fully compliant. In contrast, 12/29 (41\%) of websites that were hosted in non-local platforms and 19/44 (43\%) of websites that were not hosted in a platform were fully compliant. The relationship between e-commerce hosting platforms and compliance is statistically significant (Fisher's, df = 4, p-value $\leq$ 0.01, $\alpha_c$ = 0.33). A post-hoc analysis of the adjusted residuals with Bonferroni correction suggests that the observed count of local-hosting platforms that were classified as fully-compliant is significantly lower than expected (adjusted residual $= -4.08 < -2.77$). Conversely, the observed count of local-hosting platforms that were classified as non-compliant is significantly higher than expected (adjusted residual $= 3.77 \geq 2.77$).~\Cref{tab:hosting_compliance} in~\autorefappendix{app:results_ext} shows a breakdown of e-commerce hosting platforms in relation to compliance. 

\subsubsection{Websites' Ranking and E-Commerce Hosting Platforms.} We also looked at the relationship between websites' ranking and e-commerce hosting platforms. We find that the relationship is more salient with local hosting platforms: 11/35 (31\%) of the top-ranked websites, 10/36 (28\%) of the mid-ranked, and 6/29 (21\%) of the low-ranked websites were hosted on local e-commerce platforms. This may explain the observed pattern of high non-compliance among top-ranked websites and those on local e-commerce hosting platforms. However, the relationship between websites' ranking and e-commerce hosting platforms is not statistically significant ($\chi^2$ = 1.31, df = 4, p-value = 0.860). See~\Cref{tab:rank_locality} in~\autorefappendix{app:results_ext} for a full breakdown of websites' ranking in relation to the hosting platform. 

\subsection{RQ3: Comparing Human and LLM Answers} \label{sec:human_vs_llm}
We then analyzed potential reasons for discrepancies between our answers to the privacy policy questions and those of the LLM at a coarse-grained level (whether the item is declared: \enquote{yes}, \enquote{no}, and \enquote{other} answers). For brevity and space constraints, in what follows, we refer to the four examined privacy policy items listed in~\cref{tab:rights} by their acronyms: retention, destruction, copy, and complaint.

\subsubsection{Agreement Rates (before correcting human errors)} 
The agreement rates between our answers and the LLM before correcting the human errors that the LLM revealed (detailed below), along with the Cohen's kappa scores (to evaluate agreement beyond chance), are above 80\% in 3 out of 4 policy questions. Namely 
87/91 (96\%); $\kappa$ = 0.91 in the retention, 84/91 (92\%); $\kappa$ = 0.85 in the destruction, 53/91 (58\%); $\kappa$ = 0.31 in the copy, and 74/91 (81\%); $\kappa$ = 0.64 in the complaint answers, with overall agreement of 298/364 (82\%); $\kappa$ = 0.67. The overall Cohen's kappa score indicates a \enquote{substantial} agreement (0.61 -- 0.80), according to Landis and Koch's scale~\cite{landis77}.~\Cref{tab:disc} in~\autorefappendix{app:results_ext} shows a full breakdown of the discrepancies per question. 
\subsubsection{Reasons for Discrepancies} \label{sec:reasons}
Our analysis for the discrepancies between our answers to privacy policy coarse-grained questions and those of the LLM reveals four main themes of potential causes for discrepancies between human and LLM privacy policy answers: \begin{inparaenum}[1)]\item human error, \item LLM error, \item LLM more permissive, \item LLM more restrictive\end{inparaenum}. The first two themes were encountered in a few cases, while the remaining were exhibited in a dozen cases. In what follows, we elaborate on each theme.~\Cref{tab:themes} lists the identified themes and subthemes.~\Cref{tab:themes_freq} in~\autorefappendix{app:results_ext} shows the breakdown of discrepancies by theme and question.~\Cref{tab:themes} in~\autorefappendix{app:codebook} lists the codebook for the potential causes of discrepancies between human and LLM privacy policy answers we identified, along with the definitions and frequencies. 


\renewcommand{\arraystretch}{1.25}
\begin{table}[!t]
\centering
\small
\caption{Summary of the identified themes and subthemes of potential causes for discrepancies between human and LLM privacy policy answers. Further details are provided in~\cref{sec:reasons}.} 
\label{tab:themes}
\begin{tabular}{p{1.5cm}p{5.5cm}}
\toprule 
\makecell[l]{\textbf{Theme}\\\textbf{[exclusive]}} &
\makecell[l]{\textbf{Subtheme}\\\textbf{[non-exclusive]}} \\
\hline 
\multirow{5}{2cm}{Human error} & 
\begin{itemize}  
\item Overlay long policy. 
\item Use of certain types of text organizing techniques. 
\item Lack of or irrelevant section headings. 
\item Scattered statements for the same policy item under numerous sections.
\item Complex statements. 
\end{itemize} \\
\hline 
LLM error	& 
\begin{itemize} 
\item Mismatched views of the privacy policy text between the LLM and the human.
\end{itemize} \\
\hline 
LLM more restrictive	&  
\begin{itemize} 
\item LLM's literal interpretation.
\end{itemize} \\
\hline 
\multirow{4}{2cm}{LLM more permissive} & 
\begin{itemize} 
\item Misaligned contexts.
\item Overanalysis. 
\item Generalization.
\item LLM's literal interpretation.
\end{itemize} \\
\bottomrule
\end{tabular}
\end{table}

\par{\textbf{Human Error.}}
We identified 4 discrepancies between human and LLM answers, which we attribute to human error. We find that human errors are associated with the presence of one or more of the following factors in the relevant privacy policies. First, \textbf{overlay long policy}, such as the \num{5930}-word policy in our case, which is well above the average policy length in our dataset (\num{1943} words). Long policies can cause humans to overlook the policy item's declaration. Second, \textbf{use of certain types of text organizing techniques}, such as accordion menus, that hide detailed text until the menu item is clicked. While they reduce cognitive load, they hinder the use of the \enquote{find} feature in our case. As a result, we overlooked the declaration of a policy item within a policy divided into 26 accordion menus. Third, \textbf{lack of or irrelevant section headings.} In poorly written privacy policies, some statements are placed in an irrelevant section. For example, in one case, we overlooked the retention period declaration because it was placed in the introductory paragraph under the title \enquote{privacy policy}. Fourth, \textbf{scattered statements for the same policy item under numerous sections.} That is, when a declaration of a policy's item is scattered among multiple sections, such as the case of a retention period statement that was declared across 12 sections, each pertaining to data related to that section, e.g., \enquote{contacting our customer service}, \enquote{camera security}, etc., which led us to record the first encountered instance only. Finally, \textbf{complex statements.} This issue was exemplified in the conditional statement: \enquote{If we change our privacy policy ...} you may have your details removed from our records if you do not agree with the changes, which binds the right to destroy data to disagreement. We missed the condition and answered \enquote{yes}, while the LLM outperformed us in capturing the subtle condition and provided a more restrictive answer \enquote{other}, which is more accurate, as the website \enquote{does not state a general right to request destruction/deletion of personal data} as the LLM reasoning clarified. We corrected all human error instances before summarizing our results in~\cref{sec:results}. 

\par{\textbf{LLM Error.}} 
We classified 2 discrepancies as LLM errors, both in the complaint question. Our analysis shows that these cases arose from \textbf{mismatched views of the privacy policy text between the LLM and the human}. Both cases were because the LLM analyzed text outside the policy area (the page's footer) that contained contact details or a link to a web form. These footer details were not specified for filing complaints related to personal data, yet the LLM interpreted them as mechanisms for filing such complaints. It is worth noting that we previously identified other forms of mismatched views of the policy text and corrected them manually before running the LLM agent. Had we left them, they would likely have caused discrepancies and been classified as LLM errors. These cases include a privacy policy that displayed an empty page on the browser but contained privacy policy text inside the HTML code. In other cases, the policy text was presented in a format that our HTML text extractor could not process, such as policy text displayed in a pop-up window, rendered dynamically via a script, or embedded in a PDF file.

\par{\textbf{LLM More Restrictive.}}
We identified 40 cases in which the LLM was more restrictive than the human: 31 in the copy, 7 in the complaint, and 2 in the destruction privacy policy questions. The possible answers for the coarse-grained privacy policy questions are: \enquote{yes}, \enquote{other}, \enquote{no}. The LLM answer is more restrictive if it answered \enquote{no} or \enquote{other} for a human's \enquote{yes}, or answered \enquote{no} for a human's \enquote{other}. 

Our analysis suggests that the main reason for more restrictive LLM answers is the \textbf{LLM's literal interpretation} of the questions. Unlike humans, LLMs may lack intuition and reason more strictly in response to the provided question. This was the case in the 31 discrepancies in the question that asks about the right to obtain a copy and includes \enquote{in a readable and clear format} as per the PDPL documents. We answered \enquote{yes} for privacy policies that declared data subjects' right to obtain a copy of their personal data, with or without literally mentioning \enquote{in a readable and clear format}. This was a conscious human decision, as the essence (obtaining a copy) was declared. We treated the missing statement by adding a comment. However, the LLM answers were more restrictive than ours for these cases, with reasoning such as: \enquote{it does not specify that the copy will be provided in a readable and clear format}. 

Another 7 cases for more restrictive LLM answers, which we attribute to literal interpretation, were exhibited in the compliant questions. The LLM discarded indirect phrases indicating filing a complaint such as \enquote{If you do not agree with any of the terms and conditions outlined in the Privacy Policy} or \enquote{if you have any concerns} despite the fact they are followed by mechanisms such as an email address, with a reasoning such as: \enquote{While the policy lists certain data subject rights and provides a general contact email for inquiries or concerns, it does not describe a specific mechanism for filing complaints or objections regarding data rights or processing}. On the other hand, we interpreted soft phrases as mechanisms for filing complaints, guided by the PDPL, which is unlike other laws (e.g., GDPR), does not include an explicit \enquote{right to object} (Art. 21~\cite{gdpr25}) that requires clear and specific statements to be listed as a right.

Finally, we identified 2 discrepancy cases in the destruction question, which we attributed to literal interpretation. The LLM answered the question with \enquote{no} in two policies that granted deletion of the account data, with reasoning indicating that the policy \enquote{does not state a right to request destruction/deletion of personal data}, while we coded these cases more permissively as \enquote{other}, acknowledging that the destruction right is declared for a specific type of data, and not as a general right.

\par{\textbf{LLM More Permissive.}}
We identified 20 cases where the LLM was more permissive than the human: 2 in the retention, 3 in the destruction, 7 in the copy, and 8 in the complaint. The possible answers for the coarse-grained privacy policy questions are: \enquote{yes}, \enquote{other}, \enquote{no}. The LLM answer is more permissive if it answered \enquote{yes} or \enquote{other} for a human's \enquote{no}, or answered \enquote{yes} for a human's \enquote{other}.

The first reason for more permissive LLM answers to the privacy policy questions is attributed to \textbf{misaligned contexts.} For example, when the privacy policy contains terminology and rights from other laws. Similarly, when the privacy policy text is directed to multiple regions with conditional statements, such as \enquote{If you are located in the EEA and the UK}. We found 10 discrepancy cases between the human and LLM answers to privacy policy questions, which we attribute to misaligned contexts. Of those, 6 cases in the copy question, where the privacy policies declared the right to \enquote{access} or \enquote{portability}, but not the right to \enquote{obtaining a copy}. The PDPL's \enquote{right of access to personal data} (Art. 5) is distinct from, and does not include, the \enquote{the right to request a copy of their Personal Data} (Art. 6)~\cite{sdaia23,sdaia23_1}. In contrast, the GDPR's \enquote{right of access} grants data subjects the right to obtain \enquote{a copy of the personal data undergoing processing}~\cite{gdpr25}. Accordingly, we coded policies that declared the right to access or the portability right as \enquote{other}. In contrast, the LLM coded these cases as \enquote{yes}, leading to discrepancies which we attribute to misaligned contexts, where the LLM answers seemed to be based on a context that differed from our PDPL-based context. 

Another 4 discrepancies in the answers, which we attribute to misaligned contexts, are found in the complaint question. Unlike the GDPR, the PDPL does not have an explicit \enquote{right to object}~\cite{gdpr25,sdaia23}. Instead, the PDPL provides a guideline for declaring the complaint mechanism: \enquote{If you have any concerns, or if we do not comply with the Personal Data Protection Law, you can file a complaint to ... using one of the following channels ...}. In our manual analysis, we searched for a statement similar to the example provided in the guideline. We coded the right to object as \enquote{other} as we viewed that as another explicit right. 
In contrast, the LLM's reasoning indicates that it was looking for the \enquote{right to object} statements (similar to the GDPR), and coded declarations for the right to object as \enquote{yes}, leading to discrepancies. 

The second reason for the LLM's more permissive answers is attributed to \textbf{overanalysis}, which we observed in 2 cases in the complaint question, where the LLM connected disfragmented statements and provided a more permissive answer. For example, when the privacy policy states that they use the data to respond to questions, complaints, and queries. Then, in another section, they provide their general contact details (not clearly specified for receiving complaints). The LLM's answers to the question regarding the declaration of complaint mechanisms were more permissive than ours, with reasoning such as \enquote{The policy provides contact mechanisms ... to exercise data rights, including withdrawing consent, and states they respond to inquiries and complaints}, whereas our answer was \enquote{no} in these cases, leading to discrepancies. 

The third reason is \textbf{generalization}, exhibited in 5 cases. Two cases in the complaint question in which the LLM considered objections about receiving marketing materials and the method for unsubscribing from them as mechanisms for filing a complaint. Similarly, 2 cases in the destruction question concern the destruction of an account or profile data. The LLM classified them as \enquote{yes}, whereas we coded them as \enquote{other} because they pertain to a specific data type, resulting in discrepancies. One additional discrepancy attributed to generalization is found in the retention question, where the LLM generalized a retention declaration for a specific data type (comments) to \enquote{yes}, whereas we coded it as \enquote{other}.  

Finally, \textbf{LLM's literal interpretation} also applies here as a reason for permissive LLM answers. We identified 3 cases of discrepancies in answers for a privacy policy that consisted solely of headings (no body content). Although the headings do not provide any meaningful information about the policy items we were asking about, the LLM coded them as \enquote{other} because of keyword matching between the privacy policy heading texts and the retention, copy, and destruction questions, but they were meaningless to human analysts who answered the questions with \enquote{no} answers. 

\subsubsection{LLM Performance (after correcting human errors)} 
\begin{table}[h!]
\small
\centering
\caption{LLM performance metrics compared with human-labeled ground truth. The ground truth is represented by the answers to the privacy policy questions by two experienced researchers after they reached complete agreement and after correcting human coding errors revealed by the LLM answers after completing the comparative analysis of the human and LLM discrepant answers.}
\label{tab:llm_scores}
\begin{tabular}{p{1cm}p{1cm}p{1.2cm}p{1cm}p{1cm}l}
\toprule
\multirow{2}{*}{Question} & \multirow{2}{*}{Accuracy}  & Macro & Macro   & Macro   & Cohen's \\
 & & Precision & Recall & F1 & Kappa \\
\hline 
      Q1   & 0.98      & 0.66       & 0.66     & 0.66    & 0.96 \\
      Q2   & 0.95      & 0.80       & 0.72     & 0.74    & 0.90 \\ 
      Q3   & 0.58      & 0.45       & 0.46     & 0.44    & 0.31 \\
      Q4   & 0.81      & 0.71       & 0.66     & 0.68    & 0.64 \\
\hline 
\makecell{Overall} & 0.83    & 0.63         & 0.63     & 0.63    & 0.69 \\
\bottomrule
\end{tabular}
\end{table}

\cref{tab:llm_scores} shows the LLM performance metrics for LLM-based privacy policy analysis for a dataset that consists of Arabic and English language privacy policy text (58/91 (64\%) Arabic and 33/91 (36\%) English). We compare LLM answers to human answers (after correcting human errors revealed through the manual inspections of the discrepancies between the human and LLM answers). The overall scores are promising. Q3 has the lowest scores. Our manual comparative analysis of the discrepancies between LLM and human answers to the privacy policy questions suggests two main reasons that contributed to the large number of discrepant answers in the copy question: \textbf{LLM's literal interpretation} of the questions (when the LLM was more restrictive than the human) and \textbf{misaligned context} (when the LLM was more permissive). A detailed explanation for these two phenomena is provided in~\cref{sec:reasons}. 

Both the performance metrics and the comparative analysis that inspected the potential reasons for discrepancies can be useful sources for future benchmarks that aim to improve automated privacy policy analysis.

\section{Summary of Results and Discussion} \label{sec:discussion}
In this section, we summarize and discuss our results.~\cref{tab:summary_results} provides a short summary of the main results. We start the discussion with a summary of the compliance assessment~(\cref{sec:discuss_compliance}) followed by a discussion of potential causes and mitigations~(\cref{sec:discuss_reasons}). We then distill a list of considerations we propose for LLM-based privacy policy analysis, which can inform future benchmarks aimed at improving LLMs for privacy policy analysis~(\cref{sec:discuss_llm}).

\begin{table*}
\centering
\caption{Summary of key findings for each research question.}
\label{tab:summary_results}
\small
\begin{tabular}{p{5cm}p{11cm}}
\toprule
Research Question & Key findings \\
\hline 
\textbf{RQ1:} How compliant are e-commerce websites operating in Saudi Arabia with the PDPL's privacy policy requirements?  &  \begin{enumerate}[1)] \item Saudi e-commerce websites remain far from compliant with the PDPL's privacy policy requirements: 9\% of websites failed to provide a privacy policy, 31\% provided privacy policies that failed to declare any of the four examined items, 29\% declared some, but not all of the items, and only 31\% declared all of them. 
\item Even when the examined privacy policy items are declared, a considerable fraction of privacy policies omit required fine-grained details.
\end{enumerate} \\
\hline  
\textbf{RQ2:} How do websites' ranking and e-commerce hosting platforms relate to compliance with the PDPL's privacy policy requirements in e-commerce websites operating in Saudi Arabia? & \begin{enumerate}[1)] \item Top-ranked websites (based on search results order) and those hosted in local e-commerce platforms tend to have higher non-compliance rates compared to mid- and low-ranked websites and those not hosted on local e-commerce platforms. 
\item The difference is statistically significant, with 60\% of the top-ranked websites and 70\% of the websites hosted in local e-commerce platforms being categorized as non-compliant.\end{enumerate} \\
\hline 
\textbf{RQ3:} How reliable are LLMs when used as an automated analysis tool for assessing privacy policy compliance with the PDPL requirements compared to human analysis? &  \begin{enumerate}[1)] \item Our comparative analysis of human and LLM answers to coarse-grained privacy policy questions shows agreement rates of over 80\% on 3 of 4 questions. The LLM robustness metrics show promising results that have the potential to improve in future benchmarks by considering several issues identified by our manual comparative analysis of human and LLM answers to privacy policy questions. 
\item Our comparative analysis shows that the LLM outperformed human accuracy in some cases, particularly in finding policy items' declarations in overly long privacy policies and in those that lack sectioning or use irrelevant section headings. 
\item Based on our analysis, we suggest considerations for LLM-based privacy policy analysis, which can inform future benchmarks, including aligning HTML inner and outer privacy policy content and specifying the boundaries of the privacy policy text by website developers, which can inform future benchmarks aimed at improving LLMs for privacy policy analysis.\end{enumerate} \\
\bottomrule
\end{tabular}
\end{table*}

\subsection{E-Commerce Websites Showed a High Non-Compliance Rate With the PDPL} \label{sec:discuss_compliance}
Our analysis of 100 e-commerce websites shows that, as of this writing, they are far from compliant with the PDPL's privacy policy requirements, across multiple levels. 

At the level of privacy policy existence, 9 websites failed to provide a privacy policy. Even when a privacy policy was made available, further analysis of the policies' content, examining four items, namely: data retention, the right to request the destruction and a copy of personal data, and a mechanism for filing complaints, revealed high non-compliance rates. Our coarse-grained analysis, in which we examined the items' declaration only, shows that across the 91 policies, the declaration rate for any of the examined items (considering both \enquote{yes} and \enquote{other} answers) ranged from 40\% to 56\% only. 

Even the policies that declared the examined items, when analyzed at a fine-grained level, often lacked required details. For example, on both the right to destruction and obtaining a copy of personal data, nearly a third of the policies were missing details about the method of exercising the right, and most of the policies (82\% and 86\%, respectively) did not provide a time frame for processing the requests. On the complaint mechanism, 67\% of the policies omitted naming the designated department and/or the officer's name, and the majority (89\%) did not specify a time frame for responding.  

Overall, out of the 100 websites we analyzed, only 31 (31\%) were fully compliant (declared all four examined privacy policy items), 29\% were partially compliant, while 40\% were non-compliant (did not declare any of the examined items or did not provide a privacy policy).

\subsection{Non-Compliance: Potential Causes and Mitigations} \label{sec:discuss_reasons}
In this section, we extend our discussion regarding websites' compliance. We discuss potential causes and mitigations for high non-compliance rates. It is worth noting that it is evident that the local authority (SDAIA) has been doing efforts towards successful implementation of the PDPL, such as launching public awareness campaigns~\cite{spa_24,ad22,ad23,ad24}, conducting workshops for stakeholders~\cite{spa_25}, and offering a bundle of supporting services and tools to individuals and businesses through the National Data Governance Platform (NDGP)~\cite{dgp25}, a sub-entity of SDAIA. As an example, the \enquote{complaints service} to file complaints related to PDPL and the \enquote{privacy impact assessment}, a tool for data controllers to assess the impact of processing personal data, identify the risks of processing, and take appropriate measures to reduce the risks. However, our results suggest that PDPL implementation has not matured enough yet, which renders many e-commerce websites non-compliant. In what follows, we discuss potential causes and mitigations for non-compliance with privacy laws and regulations.

\subsubsection{Lax Enforcement}
The declaration of privacy laws and regulations by the designated authority is not sufficient for implementation. It must be accompanied by effective enforcement, including continuous monitoring, reporting mechanisms, and penalties for non-compliant controllers. However, before activating monitoring, reporting, and penalties, effective awareness and support must be provided, particularly for small and medium-sized businesses. 

\subsubsection{Need for Carefully Designed Awareness for Both Data Subjects and Controllers}
We lack data on the level of awareness among data subjects and data controllers regarding the PDPL in Saudi Arabia. However, we draw on previous work in other countries. First, from data controllers' perspective, we find that lack of awareness among data controllers, especially in developing countries and small and medium-sized businesses, was reported as an obstacle to businesses subject to GDPR~\cite{agbali20,sirur18,Li19}. These studies suggest the need for state-issued guidance, training, industry-wide awareness, and consideration of the multiple stakeholders impacted by the PDPL within the organization~\cite{sirur18,agbali20}. Second, from the data subjects' perspective, prior work provides mixed findings. For example, a study in Greece concluded that data subjects lack knowledge of their GDPR rights and that the source of knowledge affects the level of knowledge, with personal interest and the Internet as the two sources associated with the greatest extent of knowledge~\cite{sideri20}. However, another study in the Netherlands found that data subjects have high awareness of the GDPR and knowledge of their rights, likely due to national awareness campaigns~\cite{strycharz20}. 

However, care needs to be taken when designing and delivering privacy awareness campaigns for data subjects. Prince et al.'s study on online privacy literacy among EU users suggests that \enquote{procedural knowledge} (the skills or know-how), such as changing privacy settings, is positively related to users' privacy empowerment, whereas \enquote{declarative knowledge}, such as awareness of the GDPR, is only partially related to it~\cite {prince24}. Shen et al.'s study, which assessed privacy awareness campaigns released by leading technology companies, found that short videos were effective at raising awareness of privacy features but not at educating viewers on how to use them~\cite{carol24}. 

Thus, we suggest that awareness should inform data subjects not only about their rights and how to ensure that websites guarantee them in their privacy policies, but also how to practice them (e.g., adjust privacy settings and cookie consent banners, and delete the user account and its data), and how to act if their rights are not guaranteed or breached. This may require more hands-on educational methods than short awareness video campaigns. Perhaps, by considering incorporating a more hands-on educational approach that teaches (e.g., in a simulated environment) how to practice privacy rights in schools' and/or universities' information technology introductory courses, or the like.


\subsubsection{Need for Continuous Support for Both Data Subjects and Controllers}
Another potential reason for the high non-compliance rate among websites is the lack of continuous support for both data subjects and controllers. From a controller perspective, the literature has proposed a variety of solutions to support data controllers: privacy policy generators~\cite{nishant,zimmeck21}, automated analysis tools to check policy compliance with the law~\cite{andow20}, and tools to detect contradicting statements in the policy~\cite{andow19}. Additionally, data controllers may also need legal advice and question-answering tools~\cite{kuchinke16,harkous18}. We lack data on the effectiveness of the supporting tools provided by the local authority~\cite{dgp25} to data controllers from the controllers' perspective. 

In terms of support for data subjects, research has proposed several approaches: privacy policy summarization~\cite{harkous18}, question-answering tools~\cite{harkous18,ravichander19,sun25}, and privacy rating visualization tools~\cite{barth21}. However, most of these solutions have not matured enough to gain widespread adoption. They also need to be usable and accessible. 

\subsubsection{Lost Accountability for Websites Hosted on E-Commerce Platforms}
Our results show that most websites hosted on local e-commerce platforms are non-compliant. One potential reason for this phenomenon is a loss of accountability. That is, website owners assume that the e-commerce hosting platform is fully responsible for consumers' data privacy and vice versa. There is no explicit mention of this case in the PDPL documents~\cite{sdaia23,sdaia23_1,sdaia24}. However, by definition, the data controller (the website owner) is the entity that \enquote{specifies the purpose and manner of Processing Personal Data, whether the data is processed by that Controller or by the Processor}~\cite{sdaia23}. The hosting platform can be considered a processor, as it is the entity that \enquote{processes Personal Data for the benefit and on behalf of the Controller}~\cite{sdaia23}. Accordingly, data controllers are the primary responsible entities for privacy policies, data practices, and subjects' rights. It is also observed that websites hosted on local e-commerce platforms tend to be for small and medium-sized businesses and freelancers, including a specific category of freelancing license in Saudi Arabia known as \enquote{productive families} (\enquote{\ar{المنتجة} \ar{الأسر}})~\cite{freelance25}, for businesses running from homes.

Thus, we suggest that non-compliance in these websites is likely due to a lack of awareness and lost accountability. E-commerce hosting platforms can play an effective role in educating, supporting, and even requiring a PDPL-compliant privacy policy.

\subsection{Considerations for Using LLMs in Privacy Policy Analysis} \label{sec:discuss_llm}
Our results show that the agreement rate between the LLM and human answers shows promise for LLMs in automating privacy policy analysis and suggests avenues for improvement by considering the identified potential reasons for the discrepancies. In what follows, we distill lessons learned from using LLMs in privacy policy analysis and suggest the following to improve LLM-based privacy policy analysis.

\subsubsection{For Website Developers}
\paragraph{\textbf{Align Inner and Outer Privacy Policy Content in HTML Pages.}}
This is based on a discrepancy case we encountered in which the HTML page had inner HTML content containing privacy policy text, whereas the outer HTML page contained no text. As a result, the human answers to the privacy policy questions differed from the LLM's due to mismatched views. Other reasons for mismatched views between the LLM and humans are plausible. For example, a privacy policy text commented out in the HTML code, which is visible to the LLM through the HTML extracted text but not to humans. Mori et al. also noted that the use of tabular format and external links in privacy policies hindered LLM performance~\cite{mori25}. We also suggest that non-textual presentations of privacy policies, such as graphical representations, may also lead to misaligned inner-outer HTML content.

\paragraph{\textbf{Specify the Policy Text Boundaries.}}
Multiple cases of discrepancies we encountered were due to the LLM considering text outside the policy text, such as the web page's header and footer, which may contain general contact details that are not explicitly declared for data rights and practices. 

\subsubsection{For Privacy Policy Analysts}
\paragraph{\textbf{Simple HTML Text Extractor Does Not Work for All.}}
We encountered multiple cases of privacy policy text that was not presented in simple HTML format. Some policies were in PDF files, pop-up windows, or were rendered dynamically through a script. These policies could not be captured statically by a simple HTML extractor like ours. Although we have not encountered graphical presentations of policies, non-text-based methods, such as privacy labels~\cite{kelley09}, icons~\cite{holtz11}, grids~\cite{zhang24}, and graphical representations, are not ideal for LLM text-based analysis. 

\paragraph{\textbf{Potential for Improvement}}
Revisiting two studies on LLM-based privacy policy analysis in a non-Western context, Mori et al.~\cite{mori25} reported that the highest F1-score obtained on LLM-based analysis of 19 Android mobile apps' privacy policies in the Japanese language using the GPT model was 0.57 with 0.02 FPR (false positive rate) and 0.51 FNR (false negative rate). Octora et al.'s~\cite{octora25} evaluation across macro-averaged metrics, achieving 0.67 precision, 0.51 recall, and 0.52 F1-score on Preplixity. Our LLM-based analysis achieved an overall performance of 0.83 accuracy, 0.63 macro precision, 0.63 macro recall, and 0.63 macro F1-score. Our LLM performance metrics represent a first attempt, using a single model and a single question format. With the insights this study revealed into the causes of the discrepancies, which ultimately affect LLM performance, the LLM's performance in evaluating a mixture of Arabic and English language privacy policies is promising and has potential for improvement. We leave testing multiple models, different question formats, and fine-grained questions to future work.

\paragraph{\textbf{Caution in Writing Privacy Policy Questions for LLM}}
Extra care is required when drafting privacy policy questions for LLMs. Unlike humans, who may make assumptions based on intuition and norms, LLMs can be literal in their interpretation of questions. This was exhibited in many discrepancy cases due to the policy question about the right to obtain a copy of personal data, which included the phrase \enquote{in a clear and readable format}. The human analysts noted the absence of the phrase in the comments section of the analysis survey and answered \enquote{yes} for the declaration of the copy data subject right, while the LLM was more restrictive and answered \enquote{no} in most cases and \enquote{others} in some.  

\subsubsection{For Privacy Policy Authors}
\paragraph{\textbf{Respect Regional Privacy Laws and Regulation Terminology.}}
Another potential source of discrepancy between the LLM and our answers was the use of policies derived from other contexts or from multiple contexts, rather than from the regional context. For example, the GDPR's \enquote{right to access} includes the right to request a copy of personal data, whereas this is not the case under the Saudi PDPL, as the two rights are separate. Similarly, the GDPR explicitly provides a \enquote{right to object}, whereas the Saudi PDPL does so implicitly. Some conditional statements misled the LLM as they are applicable to citizens of Europe or the UK, but not to those of Saudi Arabia. Mixed-context policies led the LLM to be more permissive in our case, but they can also yield more restrictive answers.

\section{Conclusion} \label{sec:conclusion}
In this paper, we analyzed 100 e-commerce websites operating in Saudi Arabia against compliance with key PDPL rights and practices at the privacy policy level. Our results showed that, as of this writing, most websites are not yet compliant with the PDPL. Additionally, our results indicated that top-ranked websites (based on search results order) and those hosted on local e-commerce platforms have the highest non-compliance rates, compared with mid- and low-ranking websites and those not hosted on any platform. Finally, we explored and assessed LLM-based privacy policy analysis. Our evaluation of the LLM compared to human answers shows promise for LLMs in automating privacy policy analysis and suggests avenues for improvement by considering the potential reasons for the identified discrepancies. We discussed areas for potential improvement and proposed evidence-based considerations to enhance LLM-based automated privacy policy analysis. Future work can evaluate LLM-based analysis of fine-grained privacy policy questions and examine multiple models.
\section{Acknowledgment} \label{sec:ack}
The authors acknowledge using AI-assisted tools as follows: \begin{inparaenum}[1)]\item We used Grammarly~\cite{grammar26}, to assist in English-language proofreading, including grammar, spelling, punctuation, and clarity. The paper is entirely written by the authors. Any corrections made by the tool were thoroughly reviewed by the authors. \item We used ChatGPT (GPT versions 4.1 and 5.5 )~\cite{chatgpt26} to assist with coding the Python scripts used in the~\cref{sec:auto_analysis}, including generating code syntax, debugging, and refinement. The concepts and logic of the code are generated by the authors. Any AI-generated code was manually reviewed and tested by the authors. 

The authors gratefully acknowledge receiving financial support by the initiative of the Sons of Abdurahman Alagil Charitable Foundation (Jarir Investment) under the sponsorship agreement with King Abdulaziz University. This support enabled participation in Symposium On Usable Privacy and Security (SOUPS), 2026, and contributed to the research development presented in this paper.
\end{inparaenum}

\bibliographystyle{ACM-Reference-Format}
\bibliography{refs}
\clearpage
\appendix
\makeatletter
\renewcommand{\thesection}{Appendix \Alph{section}}
\makeatother
\onecolumn
\section{Extended Methods} \label{app:method_ext}
In this section, we provide additional methodological details.

\begin{table*}[h!]
\small
\centering
\caption{Supporting clauses for the analyzed items derived from the PDPL official documents: law~\cite{sdaia23}, implementing regulation~\cite{sdaia23_1}, and privacy policy guideline~\cite{sdaia24}.}
\label{tab:rights_ext}
\resizebox{0.9\textwidth}{!}{
\begin{tabular}{p{2cm}p{15cm}}
\toprule
Policy item & PDPL text  \\
\hline 
Privacy policy & 
\begin{enumerate} 
\item \enquote{The Controller shall \textbf{use a privacy policy and make it available to Data Subjects} for their information prior to collecting their Personal Data} (Law, Art. 12)~\cite{sdaia23}.
\item \enquote{The Controller shall provide access to the Privacy Policy and ensure that its content is written in clear, non-misleading, easy-to-read, and \textbf{understandable language suitable for the comprehension level of all categories of Data Subjects}} (Guideline)~\cite{sdaia24}.
\item \enquote{The Privacy Policy shall be \textbf{designed in a language suitable for the target audience}} (Guideline)~\cite{sdaia24}.
\item \enquote{\textbf{Adding other related links to the Privacy Policy}, such as Terms \& Conditions, Cookie Policy, and Personal Data Protection Law}.(Guideline)~\cite{sdaia24}.
\end{enumerate} \\
\hline

Retention & 
\begin{enumerate} 
\item \enquote{the Controller shall maintain records ... The records shall contain the following
information at a minimum: ... \textbf{6-The expected period for which Personal Data shall be retained.}} (Law Art. 31)~\cite{sdaia23}.
\item \enquote{... inform the Data Subject of the following: ... d) \textbf{The period for which the Personal Data will be stored, or if that is not possible, the criteria used to determine that period.}} (Reg., Art. 4)~\cite{sdaia23_1}. 
\item \enquote{The record of Personal Data Processing activities shall include, at a minimum, the following: ... e) \textbf{Retention periods for each Personal Data category, where possible}} (Reg., Art. 33)~\cite{sdaia23_1}. 
\end{enumerate}  \\
\hline 

Destruction & 
\begin{enumerate} 
\item \enquote{Data Subject shall have the following rights ... \textbf{5-The right to request a Destruction of their Personal Data held by the Controller}} (Law, Art. 4)~\cite{sdaia23}. 
\item \enquote{\textbf{1- The Controller shall destroy Personal Data} in any of the following cases: a) Upon Data Subject's request} (Reg., Art. 8)~\cite{sdaia23_1}. \item \enquote{Eighth: Personal Data Subjects Rights ... 5. \textbf{The Right to request the destruction of Personal Data held by the Controller}} (Guidelines)~\cite{sdaia24}. 
\item \enquote{Eighth: Personal Data Subjects Rights ... The Controller shall clarify the rights of Data Subjects ... \textbf{along with the method of exercising these rights.} The Controller shall also \textbf{provide appropriate communication channels to respond to Data Subjects' requests} ... These channels may include: \textbf{(emails, text messages, communication via electronic applications).} The Controller shall also \textbf{specify the time taken for response}} (Guidelines)~\cite{sdaia24}. \item \enquote{The Controller shall also \textbf{clarify methods used to destroy Personal Data after its intended purpose is fulfilled}, ensuring that it cannot be viewed or recovered.} (Guideline)~\cite{sdaia24}.
\end{enumerate} \\
\hline 

Copy & 
\begin{enumerate} 
\item \enquote{Data Subject shall have the following rights ... \textbf{3-The right to request obtaining their Personal Data held by the Controller in a readable and clear format}} (Law, Art. 4)~\cite{sdaia23}. 
\item \enquote{Data Subject shall have the right to \textbf{request a copy of their Personal Data in a readable and clear format}, subject to the following: ... 2- Personal Data shall be provided to the Data Subject \textbf{in a commonly used electronic format and the Data Subject may request a printed hard copy if feasible}} (Reg., Art. 6)~\cite{sdaia23_1}. 

\item \enquote{Eighth: Personal Data Subjects Rights ... The Controller shall clarify the rights of Data Subjects ... \textbf{along with the method of exercising these rights.} The Controller shall also \textbf{provide appropriate communication channels to respond to Data Subjects' requests} ... These channels may include: \textbf{(emails, text messages, communication via electronic applications).} The Controller shall also \textbf{specify the time taken for response}} (Guidelines)~\cite{sdaia24}. 

\item \enquote{Eighth: Personal Data Subjects Rights ... \textbf{3. The Right to request access to Personal Data held by the Controller in a readable and clear format} ... whether such Personal Data is \textbf{in a commonly used format if feasible, or providing a printed hard copy of such data}} (Guidelines)~\cite{sdaia24}.
\end{enumerate} \\
\hline 

Complaint & 
\begin{enumerate} 
\item \enquote{\textbf{A Data Subject may submit to the Competent Authority any complaint} that may arise out of the implementation of this Law and the Regulations. \textbf{The Regulations shall set out the rules for processing the complaints that may arise from implementing this Law and the Regulations.}} (Law, Art. 34).

\item \enquote{3- The personal data protection officer is responsible for monitoring the implementation of the provisions of the Law and its Regulations ... \textbf{and receiving requests related to Personal Data} ... Specifically, their responsibilities include: ... \textbf{e) Responding to requests from Data Subjects and addressing complaints filed by them}}. (Reg., Art. 32).

\item \enquote{Eighth: Personal Data Subjects Rights ... \textbf{7. The Right to submit any complaint related to applying the provisions of the Law to the Competent Authority}} (Guidelines)~\cite{sdaia24}.

\item \enquote{Ninth: Complaint and Objection Filing Mechanism ... \textbf{The Controller shall provide a mechanism for filing complaints and objections} ... such as failure to enable them to exercise their rights related to the processing of Personal Data ... or if there are objections to processing. This can be made by determining \textbf{the name of the department or division involved in receiving and processing complaints, its contact details, and the period specified for processing such complaints, in addition to providing information about the Competent Authority in the event of Data Subjects' dissatisfaction with the results of complaint processing by the Controller}} (Guidelines)~\cite{sdaia24}.
\end{enumerate} \\

\bottomrule
\end{tabular}
} 
\end{table*}

\begin{table}[h!]
\small
\centering
\caption{Distribution of the 100 randomly selected domains ranking across their original lists (Arabic and English), divided into three thirds (top, mid, and low). We randomly selected 100 websites for our analysis (63 out of 102 domains from the Arabic list and 37 out of 59 domains from the English list).}
\label{tab:search_rank}

\begin{tabular}{llll}
\toprule
Rank  & Ar  & En  & Ar \& En \\
\hline 
Top-rank &   22  & 13  & 35 \\ 
\hline 
Mid-rank &   22 & 14 & 36 \\ 
\hline 
Low-rank &  19  & 10 & 29 \\ 
\hline 
Total & 63 & 37  & 100 \\ 
\bottomrule
\end{tabular}
\end{table}

\begin{table*}[h!]
\small
\centering
\caption{Distribution of the e-commerce hosting services identified in the websites included in our analysis.}
\label{tab:hosting_srvs}
\begin{threeparttable}

\begin{tabular}{p{2cm}p{0.5cm}p{5cm}p{2cm}p{0.5cm}}
\toprule
Locality & \# & Hosting service &	\makecell[l]{Country\\of origin} & \#\\
\hline
 \multirow{6}{2cm}{Non-local} & \multirow{6}{0.5cm}{29} & Shopify~\cite{shopify} & Canada & 9 \\
 & & Magento~\cite{magento} & US & 5 \\
 & & Salesforce Commerce Cloud~\cite{salesforceCommerce}	& US & 5\\
 & & WooCommerce~\cite{woocommerce}	& South Africa & 6 \\
 & & Microsoft Dynamics 365 Commerce~\cite{microsoftDynamics365Commerce} & US & 1\\
 & & HCL Commerce~\cite{hclCommerce}	& India & 2 \\
 & & Amazon Webstore\tnote{1}	& US & 1 \\
\hline 
 \multirow{2}{*}{Local} & \multirow{2}{0.5cm}{27} & Salla~\cite{salla25}	& Saudi Arabia & 20 \\
 & & Zid~\cite{zid25}	& Saudi Arabia & 7\\
 \hline 
 No-hosting & 44 & n.a. & n.a. & 44 \\ 
 \hline 
Total & -- & -- & -- & 100 \\
\bottomrule  
\end{tabular}
\begin{tablenotes}
  \item[1] Amazon Webstore is deprecated~\cite{pixel26,failory26}. 
  \end{tablenotes}
\end{threeparttable}
\end{table*}

\begin{table*}[h!]
\small
\centering
\caption{Distribution of the e-commerce website according to a two-level classification: business model and product type(s). Some categories combine multiple related products. Some websites fell into multiple categories. That is, the product type(s) layer is non-exclusive, thus, the sum of percentages exceeds 100\%.}
\label{tab:categories}
\begin{tabular}{p{5cm}p{0.5cm}p{0.5cm}}
\toprule
Category & \# & \% \\
\hline 
\multicolumn{3}{l}{Business model level [exclusive]} \\
\hline 
retail       & 75 & 75\% \\
service      &	22 & 22\% \\
marketplace  &	3 & 3\% \\
\hline 
\multicolumn{3}{l}{Product type(s) level [non-exclusive]} \\
\hline 
beauty & 18 & 18\% \\
appliances/electronics/gadgets &	17 & 17\% \\ 
fashion/apparel &	16 & 16\% \\ 
sports/outdoor &	9 & 9\% \\
marketing/design/printing/branding &	7 & 7\% \\
home supplies &	7 & 7\% \\
telecom/internet & 	5 & 5\% \\
hypermarket/grocery &	4 & 4\% \\
general &	3 & 3\% \\
video games/game supplies &	3 & 3\% \\
books/media/stationary/gifts &	3 & 3\% \\
e-commerce hosting &	3 & 3\% \\
food/beverage &	2 & 2\% \\
healthcare &	2 & 2\% \\
travel/aviation/hospitality &	2 & 2\% \\
home furniture &	2 & 2\% \\
pharmacy &	1 & 1\% \\
office furniture/office supplies &	1 & 1\% \\
automotive & 	1 & 1\% \\
e-commerce intermediary &	1 & 1\% \\
toys/kids products &	1 & 1\% \\
software &	1 & 1 \% \\
philanthropy &	1 & 1 \% \\
marketing/printing/branding &	1 & 1 \% \\
kitchen furniture/kitchen supplies &	1 & 1 \% \\
\bottomrule
\end{tabular}
\end{table*}

\begin{table}[h!]
\small
\centering
\caption{Inter-Rater Reliability (IRR) between two independent human coders before they met and discussed disagreements and reached consensus on the codes. The dataset included in the IRR computation consists of websites (\textit{N = 91}) coded for four coarse-grained privacy policy questions (the first nine websites were excluded as they were coded jointly and coders reached complete agreement). The questions were about the privacy policy's declarations of: (1) personal data retention period (retention), (2) the right to request the destruction of personal data (destruction), (3) the right to request a copy of personal data (copy), and (4) a mechanism for filing complaints (complaint). The possible codes for each of the four examined questions are (\enquote{yes}, \enquote{no}, and \enquote{other}). We computed the IRR for the first 20\% of the dataset (Part 1, \textit{N = 18}) and for the remaining 80\% (Part 2, \textit{N = 73}). Finally, we computed IRR for the whole dataset (Overall: \textit{N = 91}).}
\label{tab:llm_metric}
\begin{tabular}{lllllllll}
\toprule 
\multirow{3}{*}{Question} & \multicolumn{8}{c}{Dataset} \\
\cline{2-9}
& \multicolumn{2}{c}{Part 1 (\textit{N = 18})} & & \multicolumn{2}{c}{Part 2 (\textit{N = 73})} & & \multicolumn{2}{c}{Overall (\textit{N = 91})}\\
\cline{2-3}\cline{5-6}\cline{8-9}
  & Agreement  & Kappa & & Agreement  & Kappa & & Agreement  & Kappa \\
\hline 
      Q1 (retention)  & 0.56       & 0.35  & &  0.89 & 0.81  & & 0.82 & 0.71\\           
      Q2 (destruction) & 0.67       & 0.51  & & 0.86  & 0.77  & & 0.82 & 0.72 \\
      Q3 (copy) & 0.67       & 0.56  & & 0.79  & 0.70  & & 0.77 & 0.67\\
      Q4  (complaint) & 0.67       & 0.48  & & 0.89  & 0.80  & & 0.85 & 0.73\\ 
\hline 
 Overall  & 0.64       & 0.50  & & 0.86  & 0.77  & & 0.82 & 0.71 \\
\bottomrule 
\end{tabular}
\end{table}
\clearpage

\section{Analysis Survey} \label{app:survey}
In this section, we provide the survey used in the manual analysis\footnote{The survey used in the study contained three additional sections: \begin{inparaenum}[1)]\item Halting the Reception of Advertising or Awareness Materials, \item Halting the Reception of Direct Marketing, and \item Communication Channels to Respond to Data Subjects' Requests Related to their Rights.\end{inparaenum} We omitted them here as we did not pursue analyzing or reporting them due to project scope adjustment.}.

\noindent \textbf{Title: Compliance Survey} \newline   
\noindent \textbf{1)} Email: \newline
\noindent[requires sign in to Google account] 

\noindent \textbf{2)} Enter password: \newline
\noindent[open-text entry]


\noindent\textbf{Part\#0: Before you begin. Setting up a clean environment (checklist)}
\hfill\\ 
\textbf{3)} Coder (analyst) name: \newline
\noindent[open-text entry] 

\noindent \textbf{4)} Are you using the Google Chrome browser's Incognito mode?
\begin{itemize}[$\circ$] 
    \item Yes
    \item No
\end{itemize}

\noindent \textbf{5)} Are you signed out of any online services from the browser?
\begin{itemize}[$\circ$]
    \item Yes
    \item No
\end{itemize}

\noindent \textbf{6)} Are all browser tabs closed?
\begin{itemize}[$\circ$]
    \item Yes
    \item No
\end{itemize}

\noindent\textbf{Part\#1: Retrieve the homepage and privacy policy content} \label{app:survey_part1}
\hfill\\ 
\noindent \textbf{7)} Enter the name of the controller as in the provided domains' list (e.g., \enquote{google.com}) \newline
\noindent[open-text entry]

\noindent \textbf{8)} In which language the controller's default homepage is presented?
\begin{itemize}[$\circ$]
    \item English 
    \item Arabic
    \item Other: [open-text entry]
\end{itemize}

\noindent \textbf{9)} Does the controller have an English homepage?
\begin{itemize}[$\circ$]
    \item Yes
    \item No
    \item Other: [open-text entry]
\end{itemize}
[If Q.9 answer is \enquote{No} go to Q.17] \newline 
\hfill\\
\noindent\textbf{[English homepage]} \newline 
\textbf{[The controller has an English homepage]} \newline 
\noindent \textbf{10)} Copy and paste the controller's English homepage URL:
\newline [open-text entry]

\noindent \textbf{11)} Retrieve the content of the controller's English homepage
\begin{itemize}[$\circ$]
    \item Done
    \item Other: [open-text entry]
\end{itemize}

\noindent \textbf{12)} On the English homepage, is there a link to the controller's English privacy policy named \enquote{privacy policy} or an equivalent name – some controllers use different terms for privacy policy (e.g. \enquote{privacy})?  \textbf{If you do not find a link in the home page, click the account or create account icons or areas, check for privacy policy there without creating an account or entering personal information.}
\begin{itemize}[$\circ$]
\item Yes, it is named \enquote{privacy policy}
\item Yes, it is named other than \enquote{privacy policy}
\item No
\item Other: [open-text entry]
\end{itemize}

\noindent[If Q.12 answer is \enquote{Yes, it is named \enquote{privacy policy}} go to Q.14. If Q.12 answer is \enquote{No} go to Q.17] \newline 
\hfill\\
\noindent\textbf{[Other names for English privacy policy]}\newline 
\noindent \textbf{13)} If the previous answer is \enquote{Yes, it is named other than \enquote{privacy policy}}, what is it named?
\newline [open-text entry]

\noindent \textbf{14)} Does the English privacy policy link work?
\begin{itemize}[$\circ$]
\item Yes
\item No
\item Other: [open-text entry]
\end{itemize}
\hfill\\
\noindent\textbf{[Retrieve English privacy policy]} \newline 
\textbf{[The controller has an English homepage AND it has a link to an English privacy policy AND the English privacy policy link works]} \newline 
\noindent \textbf{15)} Copy and paste the controller's English privacy policy URL:
\newline [open-text entry]

\noindent \textbf{16)} Retrieve the content of the controller's English privacy policy. 
\begin{itemize}[$\circ$]
\item Done
\item Other: [open-text entry]
\end{itemize}

\noindent \textbf{17)} Does the controller have an Arabic homepage?
\begin{itemize}[$\circ$]
\item Yes
\item No
\item Other: [open-text entry]
\end{itemize}

\noindent[If Q.17 answer is \enquote{No} go to Q.25] \newline 
\hfill\\
\noindent\textbf{[Arabic homepage]} \newline 
\textbf{[The controller has an Arabic homepage]} \newline 
\noindent \textbf{18)} Copy and paste the controller's Arabic homepage URL:
\newline [open-text entry]

\noindent \textbf{19)} Retrieve the content of the controller's Arabic homepage.
\begin{itemize}[$\circ$]
\item Done
\item Other: [open-text entry]
\end{itemize}

\noindent \textbf{20)} On the Arabic homepage, is there a link to the controller's Arabic privacy policy named \enquote{\ar{الخصوصية} \ar{سياسة}} or an equivalent name – some controllers use different terms for privacy policy? If you do not find a link on the home page, click the account or create account icons or areas, and check the privacy policy there without creating an account or entering personal information.
\begin{itemize}[$\circ$]
\item Yes, it is named \enquote{\ar{سياسة الخصوصية}}
\item Yes, it is named other than  \enquote{\ar{سياسة الخصوصية}}
\item No
\item Other: [open-text entry]
\end{itemize}

\noindent[If Q.20 answer is \enquote{Yes, it is named \enquote{\ar{سياسة الخصوصية}}} go to Q.22. If Q.20 answer is \enquote{No} go to Q.25] \newline 
\hfill\\
\noindent\textbf{[Other names for Arabic privacy policy]}\newline  
\noindent \textbf{21)} If the previous answer is \enquote{Yes, it is named other than \enquote{\ar{سياسة الخصوصية}}}, what is it named? \newline 
[open-text entry] \newline 
\hfill\\
\noindent\textbf{[Arabic privacy policy]}\newline
\noindent\textbf{[The controller has an Arabic homepage and it has a link to an Arabic privacy policy]} \newline 
\noindent \textbf{22)} Does the Arabic privacy policy link work?
\begin{itemize}[$\circ$]
\item Yes
\item No
\item Other: [open-text entry]
\end{itemize}

\noindent[If Q.22 answer is \enquote{Yes} or \enquote{Other}, go to Q.23. If Q.22 answer is \enquote{No} go to Q.25]\newline 
\hfill\\
\textbf{[Retrieve Arabic privacy policy]}\newline 
\textbf{[The controller has an Arabic homepage AND it has a link to an Arabic privacy policy AND the Arabic privacy policy link works]}\newline 
\noindent \textbf{23)} Copy and paste the controller's Arabic privacy policy URL:
\newline [open-text entry]

\noindent \textbf{24)} Retrieve the content of the controller's Arabic privacy policy. 
\begin{itemize}[$\circ$]
\item Done
\item Other: [open-text entry]
\end{itemize}
\hfill\\
\noindent \textbf{[Branch: either proceed or terminate]}\newline 
\noindent \textbf{25)} If you found a policy (either in English or Arabic), visit the privacy policy page of the controller's default homepage. If no policy exists (either in English or Arabic), choose \enquote{No policy exists}.
\begin{itemize}[$\circ$]
\item Done
\item No policy exists
\end{itemize}

\noindent[If Q.25 answer is \enquote{No policy exists} go to Part\#4 -- Q.58]\newline 
\hfill\\
\noindent\textbf{[Working default homepage's privacy policy]}\newline 
\noindent \textbf{26)} Enter the URL for the privacy policy page of the controller's default homepage.
\newline [open-text entry]

\noindent \textbf{27)} In which language is the privacy policy page of the controller's default homepage presented?
\begin{itemize}[$\circ$]
\item English 
\item Arabic
\item Other: [open-text entry]
\end{itemize}

\noindent\textbf{Part\#2: Privacy policy high-level presentation}\label{app:survey_part2}
\hfill\\
\noindent \textbf{28)} At this point, enter the data collection starting date as shown in your computer's clock:
\newline [Month-Day-Year]

\noindent \textbf{29)} At this point, enter the data collection starting time as shown in your computer's clock:
\newline [Time]

\noindent \textbf{30)} How is the privacy policy mainly presented?
\begin{itemize}[$\circ$] 
\item Textual
\item Graphical
\item Other: [open-text entry]
\end{itemize}

\noindent \textbf{31)} Does the privacy policy contain any of the following: (select all that apply)
\begin{itemize} [$\square$]
\item Table of contents
\item Additional link related to privacy policy, such as \enquote{terms and conditions}, \enquote{cookie policy}, and \enquote{personal data protection law}, etc
\item None of the above
\item Other: [open-text entry]
\end{itemize}

\noindent \textbf{32)} If you have additional comments or observations, please add them here.
\newline [open-text entry]

\noindent\textbf{Part\#3: Measure privacy policy compliance with PDPL}\label{app:survey_part3} 
\hfill\\
\noindent \textbf{Part\#3.1: Retention Period}

\noindent \textbf{33)} \textbf{[if policy\_version = \enquote{English}]:} Search for terms like \textbf{\protect\enquote{retent}} or similar terms in the privacy policy to find information about the retention period or the criteria used to determine it. Also, skim through the policy headings to double-check.

\noindent\textbf{[else if policy\_version = \enquote{Arabic}]:} Search for terms like \textbf{\protect\enquote{\ar{حتفظ}},
\protect\enquote{\ar{حفظ}},
\protect\enquote{\ar{حتفاظ}}} or similar terms in the privacy policy to find information about the retention period or the criteria used to determine it. Also, skim through the policy headings to double-check.
\begin{itemize}[$\circ$]
\item Done
\item Other: [open-text entry]
\end{itemize}

\noindent \textbf{34)} Does the privacy policy inform data subjects about the retention period or the criteria used to determine how long the controller will retain their personal data?
\begin{itemize}[$\circ$]
\item Yes
\item No
\item Other: [open-text entry]
\end{itemize}

\noindent \textbf{35)} Copy and paste the section heading(s) in the privacy policy that describe the retention period or the criteria used to determine how long the controller will retain the data subjects' personal data. If not specified in the privacy policy, write \enquote{n.a.} Use double quotes to mark the start and end of the copied heading(s), and leave an empty line between multiple headings.
\newline [open-text entry]

\noindent \textbf{36)} Copy and paste the paragraph(s) in the privacy policy that describe the retention period or the criteria used to determine how long the controller will retain the data subjects' personal data. If not specified in the privacy policy, write \enquote{n.a.} Use double quotes to mark the start and end of the copied paragraph(s), and leave an empty line between multiple paragraphs.
\newline [open-text entry]

\noindent \textbf{37)} If you have additional comments or observations, please add them here.
\noindent[open-text entry] \newline 

\noindent\textbf{Part\#3.2: Request Destruction of Personal Data} \newline
\noindent \textbf{38) [if policy\_version = \enquote{english}]:} Search for terms like \textbf{\protect\enquote{delet}, \protect\enquote{clos}, \protect\enquote{remov}, \protect\enquote{destruct}, \protect\enquote{eras}} or similar terms in the privacy policy to find information about destruction of personal data. Also, skim through the policy headings to double-check. \newline 
\textbf{[else if policy\_version = \enquote{arabic}]:} Search for terms like \textbf{\protect\enquote{\ar{تلف}}, \protect\enquote{\ar{تلاف}}, \protect\enquote{\ar{زال}}, \protect\enquote{\ar{غلق}}, \protect\enquote{\ar{غلاق}}, \protect\enquote{\ar{مسح}}} or similar terms in the privacy policy to find information about destruction of personal data. Also, skim through the policy headings to double-check.
\begin{itemize}[$\circ$]
\item Done
\item Other: [open-text entry]
\end{itemize}

\noindent \textbf{39)} Does the privacy policy inform data subjects about the right to request the destruction of their personal data held by the controller?
\begin{itemize}[$\circ$]
\item Yes
\item No
\item Other: [open-text entry]
\end{itemize}

\noindent \textbf{40)} Copy and paste the section heading(s) in the privacy policy that describe the data subjects' right to the destruction of their personal data held by the controller. If not specified in the privacy policy, write \enquote{n.a.} Use double quotes to mark the start and end of the copied heading(s), and leave an empty line between multiple headings.
\newline [open-text entry]

\noindent \textbf{41)} Copy and paste the paragraph(s) in the privacy policy that describe the data subjects' right to the destruction of their personal data held by the controller, \textbf{\protect\textit{\uline{including any related information such as the method of exercising the right, response time, etc.}}} If not specified in the privacy policy, write \enquote{n.a.} Use double quotes to mark the start and end of the copied paragraph(s), and leave an empty line between multiple paragraphs.\newline
[open-text entry] \newline 
\noindent \textbf{42)} If you have additional comments or observations, please add them here.
\newline [open-text entry] \newline 

\noindent \textbf{Part\#3.3: Request A Copy of Personal Data in a Readable and Clear Format} \newline 

\noindent \textbf{43) [if policy\_version = \enquote{english}]:} Search for terms like \textbf{\protect\enquote{copy}, \protect\enquote{readable}, \protect\enquote{clear}, \protect\enquote{format}} or similar terms in the privacy policy to find information about requesting a copy of data in a readable and clear format. Also, skim through the policy headings to double-check.

\noindent \textbf{[else if policy\_version = \enquote{arabic}]:} Search for terms like \textbf{\protect\enquote{\ar{صيغ}}, 
\protect\enquote{\ar{مقروء}},
\protect\enquote{\ar{واضح}},
\protect\enquote{\ar{نسخ}}} or similar terms in the privacy policy to find information about requesting a copy of data in a readable and clear format. Also, skim through the policy headings to double-check.
\begin{itemize}[$\circ$]
\item Done
\item Other: [open-text entry]
\end{itemize}

\noindent \textbf{44)} Does the privacy policy inform data subjects about the right to request a copy of their personal data in a readable and clear format?
\begin{itemize}[$\circ$]
\item Yes
\item No
\item Other: [open-text entry]
\end{itemize}

\noindent \textbf{45)} Copy and paste the section heading(s) in the privacy policy that describe the data subjects' right to request a copy of their personal data in a readable and clear format. If not specified in the privacy policy, write \enquote{n.a.} Use double quotes to mark the start and end of the copied heading(s), and leave an empty line between multiple headings.
\newline [open-text entry]

\noindent \textbf{46)} Copy and paste the paragraph(s) in the privacy policy that describe the data subjects' right to request a copy of their personal data in a readable and clear format, \textbf{\protect\textit{\uline{including any related information such as the method of exercising the right, response time, etc.}}} If not specified in the privacy policy, write \enquote{n.a.} Use double quotes to mark the start and end of the copied paragraph(s), and leave an empty line between multiple paragraphs.
\newline [open-text entry]

\noindent \textbf{47)} If you have additional comments or observations, please add them here. \newline 
\noindent[open-text entry] \newline

\noindent \textbf{Part\#3.4: Submitting Complaints or Objections Related to Data Subjects' Rights} \newline 

\noindent \textbf{53) [if policy\_version = \enquote{english}]:} Search for terms like \textbf{\protect\enquote{complain}, \protect\enquote{object}, \protect\enquote{concern}} or similar terms in the privacy policy to find information about the mechanism provided by the controller for filing complaints and objections related to their data rights and processing. Also, skim through the policy headings to double-check.

\noindent \textbf{[if policy\_version = \enquote{arabic}]:} Search for terms like  \textbf{\protect\enquote{\ar{مخاوف}}, \enquote{\protect\ar{شكوى}}, \protect\enquote{\ar{عتراض}}, \protect\enquote{\ar{شكاوى}}} or similar terms in the privacy policy to find information about the \textbf{mechanism provided by the controller for filing complaints and objections related to their data rights and processing.} Also, skim through the policy headings to double-check.
\begin{itemize}[$\circ$]
\item Done
\item Other: [open-text entry]
\end{itemize}

\noindent \textbf{54)} Does the privacy policy inform data subjects about the mechanism provided by the controller for filing complaints and objections related to their data rights and processing?
\begin{itemize}[$\circ$]
\item Yes
\item No
\item Other: [open-text entry]
\end{itemize}

\noindent \textbf{55)} Copy and paste the section heading(s) in the privacy policy that describe the mechanism provided by the controller for filing complaints and objections related to their data rights and processing. If not specified in the privacy policy, write \enquote{n.a.} Use double quotes to mark the start and end of the copied heading(s), and leave an empty line between multiple headings.
\newline [open-text entry]

\noindent \textbf{56)} Copy and paste the paragraph(s) in the privacy policy that describe the mechanism provided by the controller for filing complaints and objections related to their data rights and processing. If not specified in the privacy policy, write \enquote{n.a.} Use double quotes to mark the start and end of the copied paragraph(s), and leave an empty line between multiple paragraphs.
\newline [open-text entry]

\noindent \textbf{57)} If you have additional comments or observations, please add them here.
\newline [open-text entry]

\noindent\textbf{Part\#4: Additional comments} \label{app:survey_part4}
\hfill\\
\noindent \textbf{58)} Do you have any additional comments or observations? \newline
[open-text entry]

\noindent \textbf{59)} At this point, enter the data collection ending date as shown in your computer's clock:\newline
[Month-Day-Year]

\noindent \textbf{60)} At this point, enter the data collection ending time as shown in your computer's clock \textbf{\protect\textit{\uline{(use 24-hour format):}}}\newline
[Time] \newline 

\noindent\textbf{[End of survey]} \newline 
\uline{You reached the end of survey. Thank You!}

\clearpage

\section{LLM List of Questions} \label{app:llm-questions}
In this section, we list the questions used by our LLM agent on each privacy policy in our dataset. The LLM retrieves the question and choices from a CSV file (these questions were also listed in~\autorefappendix{app:survey} (Q.34, Q.39, Q.44, and Q.54) and were answered manually by human analysts). The answers are (\enquote{Yes}, \enquote{No}, \enquote{Other (please specify):}. The LLM-based analysis was on the coarse-grained privacy policy questions. 
\\\\
\noindent \textbf{LLM Questions} \newline   
\begin{enumerate}
    \item Does the privacy policy inform data subjects about the retention period or the criteria used to determine how long the controller will retain their personal data?
    \begin{itemize}[$\circ$]
        \item Yes
        \item No
        \item Other (please specify):
    \end{itemize}

    \item Does the privacy policy inform data subjects about the right to request the destruction of their personal data held by the controller?
        \begin{itemize}[$\circ$]
        \item Yes
        \item No
        \item Other (please specify):
    \end{itemize}
    
    \item Does the privacy policy inform data subjects about the right to request a copy of their personal data in a readable and clear format?
        \begin{itemize}[$\circ$]
        \item Yes
        \item No
        \item Other (please specify):
    \end{itemize}

    \item Does the privacy policy inform data subjects about the mechanism provided by the controller for filing complaints and objections related to their data rights and processing?
        \begin{itemize}[$\circ$]
        \item Yes
        \item No
        \item Other (please specify):
    \end{itemize}
    
\end{enumerate}

\clearpage

\onecolumn
\section{Extended Results} \label{app:results_ext}
\begin{table*}[h!]
\small
\centering
\caption{Summary of basic features for the websites included in our analysis. The last column reports statistics for the analyzed websites' default homepages (whether it is in English or Arabic).}
\label{tab:summary_ext}
\begin{threeparttable}
\begin{tabular}{lll}
\toprule
\multirow{3}{*}{Feature} &   \multicolumn{2}{c}{Default homepage}\\
    &   \multicolumn{2}{c}{(English or Arabic)}\\
    
    & \multicolumn{2}{c}{\textit{N} = 100}  \\
\hline 
$\rightarrow$  Working link to privacy policy  &  91/100 & (91\%)  \\
\hline 
$\rightarrow$$\rightarrow$ Link labeled in Arabic   &  63/91 & (69\%)  \\
$\rightarrow$$\rightarrow$ Link labeled in English  & 28/91 & (31\%)  \\
\hline 
$\rightarrow$$\rightarrow$ Direct link   &  88/91 & (97\%)   \\
$\rightarrow$$\rightarrow$ Indirect link  & 3/91 & (3\%)  \\
\hline 
$\rightarrow$$\rightarrow$ Link labeled \enquote{privacy policy}  & 59/91 & (65\%)  \\
$\rightarrow$$\rightarrow$ Link labeled other than \enquote{privacy policy}   & 32/91 & (35\%)  \\
$\rightarrow$$\rightarrow$ Link labeled without the word \enquote{privacy}\tnote{1}   & 4/91 & (4\%)  \\
\hline 
$\rightarrow$$\rightarrow$ Privacy policy content in Arabic by default  & 58/91 & (64\%)  \\
$\rightarrow$$\rightarrow$ Privacy policy content in English by default & 33/91 & (36\%)   \\
\bottomrule
\end{tabular}
\begin{tablenotes}
  \item[1] These four websites were also counted in the previous row as they are also with \enquote{Link labeled other than \enquote{privacy policy}}. 
  \end{tablenotes}
\end{threeparttable}
\end{table*}

\begin{table*}[h!]
\small
\centering
\caption{Distribution of compliant websites in relation to website rankings. The second column lists the observed counts, the expected counts, the percentage of the counts from the total instances of the column (Column \%), and the adjusted residuals (Adj. Residual). The Chi-squared test result is significant ($\chi^2$ = 11.7, df = 4, p-value = 0.019, $\alpha_c$ = 0.24). Adjusted residuals greater than 2.77 or less than $-2.77$ are marked with bold font and a superscripted star \textbf{($^*$)}, marking the cell(s) contributing to the statistically significant p-value.}
\label{tab:rank_compliance}
\begin{tabular}{p{3cm}p{2cm}p{2cm}p{2cm}p{2cm}p{1cm}}
\toprule
Compliance   & Value & Top-ranked & Mid-ranked & Low-ranked & Total \\
\hline
Fully compliant      & Observed      & 7           & 13      & 11     & 31 \\	
                     & Expected      &  10.85      & 11.16   & 8.99   &    \\	
                     & Column \% &  20\%    & 36.11\%    & 37.93\%   &    \\	
                     & Adj. Residual  & $-1.75$       & 0.83    & 0.96  &    \\	
\hline 
Partially compliant	 & Observed      &  7          & 15      &  7      & 29 \\
                     & Expected      & 10.15       & 10.44   & 8.41    &    \\	
                     & Column \% & 20\%        & 41.67\% & 24.14\%    &    \\		
                     & Adj. Residual  & $-1.46$       & 2.09    & $-0.68$  &    \\	
\hline 
Non-compliant	     & Observed      & 21          & 8      & 11      & 40 \\	
                     & Expected      & 14.00       & 14.40  &  11.60  &    \\	
                     & Column \% & 60\%        & 22.22\%   &  37.93\%   &    \\		
                     & Adj. Residual  & \textbf{3.00$^*$}  & $-2.72$ &  $-0.27$ &    \\	
\hline 
Total                & --     & 35  &    36 &    29 &   100 \\
\bottomrule
\end{tabular}
\end{table*}

\begin{table*}[h!]
\small
\centering
\caption{Distribution of compliant websites in relation to e-commerce hosting platform type. The second column lists the observed counts, the expected counts, the percentage of the counts from the total instances of the column (Column \%), and the adjusted residuals (Adj. Residual). The Chi-squared test result is significant (Fisher's test, df = 4, p-value $\leq$ 0.01, $\alpha_c$ = 0.333). Adjusted residuals greater than 2.77 or less than $-2.77$ are marked with bold font and a superscripted star \textbf{($^*$)}, marking the cell(s) contributing to the statistically significant p-value.}
\label{tab:hosting_compliance}
\begin{tabular}{p{3cm}p{2cm}p{2cm}p{2cm}p{2cm}p{1cm}}
\toprule
Compliance   & Value &  Local & Non-Local & No-Hosting & Total \\
\hline 
Fully compliant       & Observed       & 0	       & 12       & 19       & 31 \\
                     & Expected      & 8.37        & 8.99     & 13.64    &    \\	
                     & Column \% & 0.00\%      & 41.38\%  & 43.18\%  &    \\	
                     & Adj. Residual  & $\mathbf{-4.08^*}$    & 1.43     & 2.33    &    \\	
\hline 
Partially compliant    & Observed       & 8           & 11	      & 10       & 38 \\
                     & Expected      & 7.83        & 8.41     & 12.76    &    \\	
                     & Column \% & 29.63\%     & 37.93\%  & 22.73\%  &    \\	
                     & Adj. Residual  & 0.08       & 1.26      & $-1.23$    &    \\	
\hline 
Non-compliant        & Observed       & 19	       & 6	      & 15       & 31 \\
                     & Expected      & 10.80       & 11.60    & 17.60    &    \\	
                     & Column \% & 70.37\%     & 20.69\%  & 34.09\%  &    \\	
                     & Adj. Residual  & \textbf{3.77$^*$}      & $-2.52$    & $-1.07$   &    \\	
\hline 
Total                & --        & 27	       &  29      &	44       & 100 \\
\bottomrule
\end{tabular}
\end{table*}

\begin{table*}[h!]
\small
\centering
\caption{Distribution of websites' e-commerce hosting services in relation to websites' rankings. The second column lists the observed counts, the expected counts, the percentage of the counts from the total instances of the column (Column \%), and the adjusted residuals (Adj. Residual). The Chi-squared test result is not significant ($\chi^2$ = 1.31, df = 4, p-value = 0.860, $\alpha_c$ = 0.08).}
\label{tab:rank_locality}
\begin{tabular}{p{3cm}p{2cm}p{2cm}p{2cm}p{2cm}p{1cm}}
\toprule
Hosting platform & Value & Top-ranked & Mid-ranked & Low-ranked & Total \\
\hline 
Local                 &  Obs.    & 11	      &  10	     & 6        & 27 \\
                     & Expected      & 9.45       & 9.72    &  7.83    &    \\	
                     & Column \% & 31.43\%    & 27.78\% & 20.69\%     &    \\	
                     & Adj. Residual  & 0.73       & 0.13    & $-0.91$   &    \\	
\hline 
Non-local            & Obs.      & 10        &  11       & 8        & 29 \\
                     & Expected      & 10.15     &  10.44    & 8.41     &    \\	
                     & Column \% & 28.57\%   & 30.56\%   & 27.59\%  &    \\	
                     & Adj. Residual  & $-0.07$     & 0.26       & $-0.20$   &    \\	
\hline 
Non-hosting          & Obs.      & 14        &  15      &	15     &  44 \\
                     & Expected      & 15.40     & 15.84    & 12.76    &    \\	
                     & Column \% & 40.00\%   & 41.67\%  & 51.72\%  &    \\	
                     & Adj. Residual  & $-0.59$     & $-0.35$     & 0.99    &    \\	
\hline
Total               &  --        &  35       &  36      &	29     &  100   \\
\bottomrule
\end{tabular}
\end{table*}

\begin{table*}[h!]
\small
\centering
\caption{Number of discrepant cases between human and LLM answers to the coarse-grained privacy policy questions.}
\label{tab:disc}
\begin{tabular}{ccp{0.1cm}p{1.2cm}p{1.2cm}p{1.2cm}p{1.2cm}p{0.1cm}p{1.2cm}}
\toprule
 \multicolumn{2}{c}{Answer}  & &  \multicolumn{4}{c}{Question} & & \multirow{3}{*}{Total} \\
\cline{1-2}\cline{4-7}
  \multirow{2}{*}{Human} & \multirow{2}{*}{LLM} & & Q1	     & Q2 	& Q3 	& Q4   & &  \\
        &     & &   (retention) & (destruct) & (copy) & (Complaint)  & &\\        
\hline 
 no &  other & & 1	& 1	& 1	& 2 & &  5 \\
 no &  yes	& & 1	& 1	& 0	& 4 & & 6 \\
\hline 
 other&  yes  & & 2	& 2	& 6	& 4 & &  14 \\
 other &  no  & & 0	& 2	& 13 & 0 & & 15 \\
\hline 
 yes &  no   & & 0	& 0	& 6	 & 7 & & 13 \\
 yes  & other & & 0	& 1	& 12 & 0 & & 13 \\
\hline 
 \multicolumn{2}{c}{Total} & & 4 & 7 & 38 & 17 & & 66 \\
\bottomrule
\end{tabular}
\end{table*}

\begin{table*}[h!]
\small
\centering
\caption{Main themes identified as causes for discrepancies between human and LLM answers to the privacy policy questions along with their frequencies per question.}
\label{tab:themes_freq}
\begin{tabular}{lp{0.1cm}p{0.1cm}p{1.2cm}p{1.2cm}p{1.2cm}p{1.2cm}p{0.1cm}p{1.2cm}}
\toprule
\multirow{3}{*}{Question}& & &\multicolumn{4}{c}{Discrepancy reason} & \\
\cline{4-7}
   & & &\makecell[l]{Human\\error} & \makecell[l]{LLM\\error}	& \makecell[l]{LLM more\\restrictive} & \makecell[l]{LLM more\\permissive} 	 & & Total \\
\hline 
Retention      & & &  2  & 0  & 0 & 2 & & 4\\
Destruct 	   & & &  2  & 0 & 2 & 3  & & 7\\
Copy	       & & &  0  & 0 & 31 & 7 & & 38 \\
Complain       & & &  0  & 2 & 7 & 8  & & 17 \\
\hline 
Total & &  &  4 & 2  & 40 & 20 &  & 66 \\
\bottomrule
\end{tabular}
\end{table*}

\clearpage

\onecolumn
\section{Codebook} \label{app:codebook}
In this section,~\cref{tab:homepage_pres} to~\cref{tab:themes} list the codebooks we created to analyze our data, divided into tables. Each table represents the codebook for a set of related questions. All codes are in small letters only. 

{
\renewcommand{\arraystretch}{1.25}
\begin{center}
\small
\begin{longtable}{p{3.5cm}p{12cm}p{1cm}}
\caption{Codebook for websites' homepage and privacy policy presentation in English and Arabic (survey Q\#1 -- Q\#27) (\textit{N = 100}). Themes are represented in grayed rows starting with \enquote{Theme}. The first column represents the code. Exclusive codes (i.e., cannot appear with other codes in the same theme are followed by \enquote{[exclusive]} either at the column label if this applies to all codes or at the code cell if it applies to a subset of the codes. The second column contains the code definition, and the last column contains the code frequency.} \label{tab:homepage_pres} \\
\toprule 
\rowcolor{gray!20}
\multicolumn{3}{l}{\grcell{\textbf{Theme:} presentation -> english\_homepage\_availability}} \\
\hline 
\rowcolor{gray!20}
\textbf{Code} [exclusive]   & \textbf{Code Definition}   & \textbf{\#}     \\
\endfirsthead
\caption[]{(continued)}\\
\endhead
\toprule
    yes;    & The homepage is available in English.	& 73 \\
\hline 
    no;	    & The homepage is not available in English.	& 22 \\
\hline
    minimal; & The website is available in English, but the English version translates the bare minimum, i.e., only the website template, such as "Add to cart" and "Search" buttons. The rest of the content remains in the other language. Nevertheless, websites coded with this code are counted alongside those available in English (based on the language selector and template changes).  & 5 \\
\hline 
\rowcolor{gray!20}
    \multicolumn{3}{l}{\grcell{\textbf{Theme:} presentation -> arabic\_homepage\_availability}} \\
    \hline
\rowcolor{gray!20}
    \textbf{Code} [exclusive]   & \textbf{Code Definition}   & \textbf{\#}     \\
    \hline
    yes; & The homepage is available in Arabic.	& 98 \\
    \hline 
    no;	 & The homepage is not available in Arabic.	& 1 \\
    \hline 
    minimal; &	The website is available in Arabic, but the Arabic version translates the bare minimum, i.e., only the website template, such as "Add to cart" and "Search" buttons. The rest of the content remains in the other language. Nevertheless, websites coded with this code are counted alongside those available in Arabic (based on the language selector and template changes). & 1 \\ 

\hline 
\multicolumn{3}{l}{\grcell{\textbf{Theme:} presentation -> english\_privacy\_policy\_link}} \\
\hline 
\rowcolor{gray!20}
\textbf{Code} [exclusive]   & \textbf{Code Definition}   & \textbf{\#}     \\
\hline 
    yes\_pp;	& The English homepage has a privacy policy link, and it is named \enquote{privacy policy}.  &
    48 \\
\hline 
    yes\_other;	& The English homepage has a privacy policy link, and it is named other than \enquote{privacy policy}. &  27 \\
\hline
    no;	& There is no privacy policy link on the English homepage.	& 3 \\
\hline 
    n.a.	& There is no English homepage.	& 22 \\
\hline 
\multicolumn{3}{l}{\grcell{\textbf{Theme:} presentation -> arabic\_privacy\_policy\_link}} \\
\hline 
\rowcolor{gray!20}
\textbf{Code} [exclusive]   & \textbf{Code Definition}   & \textbf{\#}     \\
\hline 
    yes\_pp;	& The Arabic homepage has a privacy policy link, and it is named \enquote{\ar{الخصوصية} \ar{سياسة}}. & 56 \\
\hline 
    yes\_other;	& The Arabic homepage has a privacy policy link, and it is named other than \enquote{\ar{الخصوصية} \ar{سياسة}}.   &	34 \\
\hline 
no\_pp;	& There is no privacy policy link on the Arabic homepage & 9 \\
\hline 
    n.a.	& There is no Arabic homepage.	& 1 \\

\hline 
\multicolumn{3}{l}{\grcell{\textbf{Theme:} presentation -> english\_working\_privacy\_policy}} \\
\hline 
\rowcolor{gray!20}
\textbf{Code} [exclusive]   & \textbf{Code Definition}   & \textbf{\#}     \\
\hline 
yes;	& The English homepage's privacy policy link is working, and the policy's content is displayed in English.	& 67 \\
\hline 
arabic\_exclusive;	& The English homepage's privacy policy link is working, and the policy's content is presented exclusively in Arabic. &	8 \\ 
\hline 

arabic\_default;	& The English homepage's privacy policy link is working, and the policy's content is presented in Arabic by default, but can be changed to English manually.	& 0 \\
\hline 
no;	& The English homepage's privacy policy link is not working.	& 0 \\
\hline 
no\_pp;	& There is no privacy policy link on the English homepage. & 3 \\
\hline 
n.a. ;	& There is no English homepage. & 22 \\
\hline 
\rowcolor{gray!20}
\multicolumn{3}{l}{\textbf{Theme:} presentation -> arabic\_working\_privacy\_policy} \\
\hline 
\rowcolor{gray!20}
\textbf{Code} [exclusive]   & \textbf{Code Definition}   & \textbf{\#}     \\
\hline 
yes;	& The Arabic homepage's privacy policy link is working, and the policy's content is displayed in Arabic.	& 85 \\
\hline 
english\_exclusive;	& The Arabic homepage's privacy policy link is working, and the policy's content is presented exclusively in English.	& 1 \\
\hline 
english\_default;	& The Arabic homepage's privacy policy link is working, and the policy's content is presented in English by default, but can be changed to Arabic manually.	& 4 \\
\hline 
no;	& The Arabic homepage's privacy policy link is not working.	& 0 \\
\hline 
no\_pp;	& There is no privacy policy link on the Arabic homepage. & 9 \\
\hline 
n.a.	& There is no Arabic homepage. & 1 \\
\hline 
\rowcolor{gray!20}
\multicolumn{3}{l}{\textbf{Theme:} presentation -> default\_homepage\_language } \\
\hline 
\rowcolor{gray!20}
\textbf{Code} [exclusive]   & \textbf{Code Definition}   & \textbf{\#}     \\
\hline 
english;	& The homepage is presented in English by default.	& 29 \\
\hline 
arabic;	& The homepage is presented in Arabic by default.	& 71 \\
\hline 
n.a.	& There is no homepage in Arabic or English.	& 0 \\
\bottomrule
\end{longtable}
\end{center}
}

{
\renewcommand{\arraystretch}{1.25}
\begin{center}
\small
\begin{longtable}{p{3.5cm}p{12cm}p{1cm}}
\caption{Codebook for websites' default privacy policy's content presentation (survey Q\#28 -- Q\#32) (\textit{N = 91}). By the \enquote{default privacy policy} we refer to the privacy policy accessed from the default homepage. Themes are represented in grayed rows starting with \enquote{Theme}. The first column represents the code. Exclusive codes (i.e., cannot appear with other codes in the same theme are followed by \enquote{[exclusive]} either at the column label if this applies to all codes or at the code cell if it applies to a subset of the codes. The second column contains the code definition, and the last column contains the code frequency.} \label{tab:policy_pres} \\
\endfirsthead
\caption[]{(continued)}\\
\endhead
\toprule
\rowcolor{gray!20}
\multicolumn{3}{l}{\textbf{Theme:} presentation -> default\_privacy\_policy\_language} \\
\hline 
\rowcolor{gray!20}
\textbf{Code} [exclusive]   & \textbf{Code Definition}   & \textbf{\#}     \\
\hline 
english;	& The default privacy policy is presented in English.	& 33 \\
\hline 
arabic; & 	The default privacy policy is presented in Arabic.	& 58 \\
\hline 
\rowcolor{gray!20}
\multicolumn{3}{l}{\textbf{Theme:} presentation -> default\_privacy\_policy\_language\_vs.\_default\_homepage\_language} \\
\hline 
\rowcolor{gray!20}
\textbf{Code} [exclusive]    & \textbf{Code Definition}   & \textbf{\#}     \\
\hline 
same;	& The default privacy policy's language is the same as the default homepage's language.	& 86 \\
\hline 
diff;	& The default privacy policy's language is different from the default homepage's language.	& 5 \\
\hline
\rowcolor{gray!20}
\multicolumn{3}{l}{\textbf{Theme:} presentation -> default\_privacy\_policy\_link} \\
\hline 
\rowcolor{gray!20}
\textbf{Code} [exclusive]    & \textbf{Code Definition}   & \textbf{\#}     \\
\hline 
direct;	& The privacy policy link is directly placed on the homepage. &	88 \\
\hline 
indirect;	& The privacy policy link is indirect and requires multiple clicks from the homepage. &	3 \\
\hline 
\rowcolor{gray!20}
\multicolumn{3}{l}{\textbf{Theme:} presentation -> default\_privacy\_policy\_textual\_graphical\_presentation} \\
\hline 
\rowcolor{gray!20}
\textbf{Code} [exclusive]    & \textbf{Code Definition}   & \textbf{\#}     \\
\hline 
textual;	& The privacy policy is mostly presented in a textual format.	& 91 \\ 
\hline 
graphical;	& The privacy policy is mostly presented in a graphical format.	& 0 \\ 
\hline 
\rowcolor{gray!20}
\multicolumn{3}{l}{\textbf{Theme:} presentation -> privacy\_policy\_content} \\
\hline 
\rowcolor{gray!20}
\textbf{Code}   & \textbf{Code Definition}   & \textbf{\#}     \\
\hline 
table\_content;	& The privacy policy contains a table of contents section at the beginning of the policy. It does not need to be clickable. It should cover all important sections of the policy.	& 10\\
\hline 
links; & The privacy policy contains additional informational links to the policy or links to organize the policy into smaller sections, such as cookie policy, personal data protection law, etc. Links that open forms where data subjects can request their rights, or links to the user account page, are not counted as additional links.	& 28 \\
\hline 
accordion\_menu;	& The privacy policy contains a working accordion menu despite its default status (collapsed or not).	& 4 \\
\hline 
none; [exclusive] 	& The privacy policy does not contain any of the above items.	& 58 \\
\hline 

\rowcolor{gray!20}
\multicolumn{3}{l}{\textbf{Theme:} presentation -> privacy\_policy\_relevance} \\
\hline 
\rowcolor{gray!20}
\textbf{Code}   & \textbf{Code Definition}   & \textbf{\#}     \\
\hline 
combined;	& The privacy policy is combined with other policies (or content) such as \enquote{terms of use} or other not directly related policies. This code does not apply when the privacy policy is combined with a closely related policy, such as a cookie policy or a security policy.& 19 \\ 
\hline 
item;	& The privacy policy was presented as an item (in a line or a few lines) inside a different page.	& 4 \\
\hline 
irrelevant;	& The privacy policy content is irrelevant to privacy, e.g., \enquote{return policy}, or is meaningless content, such as (in progress). 	& 4 \\
\hline 
none; [exclusive] & The privacy policy does not contain any of the above items.	& 68 \\
\bottomrule
\end{longtable}
\end{center}
}

{
\renewcommand{\arraystretch}{1.25}
\begin{center}
\small
\begin{longtable}{p{3.5cm}p{12cm}p{1cm}}
\caption{Codebook for the personal data retention period in the privacy policy text (survey Q\#33 -- Q\#37) (\textit{N = 91}). Themes are represented in grayed rows starting with \enquote{Theme}. The first column represents the code. Exclusive codes (i.e., cannot appear with other codes in the same theme are followed by \enquote{[exclusive]} either at the column label if this applies to all codes or at the code cell if it applies to a subset of the codes. The second column contains the code definition, and the last column contains the code frequency.} \label{tab:retention} \\
\endfirsthead
\caption[]{(continued)}\\
\endhead
\toprule
\rowcolor{gray!20}
\multicolumn{3}{l}{\textbf{Theme:} retention -> declaration} \\
\hline 
\rowcolor{gray!20}
\textbf{Code} [exclusive]   & \textbf{Code Definition}   & \textbf{\#}     \\
\hline 
yes;	& The privacy policy declares the retention period or the criteria to determine it.	& 41 \\
\hline 
no;	& The privacy policy does not declare the retention period or the criteria to determine it.	& 49 \\
\hline 
other;	& The privacy policy declares the retention period or the criteria to determine it, but only for specific cases (see the \enquote{other} sub-themes below).	& 1 \\

\hline 
\rowcolor{gray!20}
\multicolumn{3}{l}{\textbf{Theme:} retention -> declaration -> \enquote{other;} -> reason for \enquote{other;}} \\
\hline 
\rowcolor{gray!20}
\textbf{Code}   & \textbf{Code Definition}   & \textbf{\#}     \\
\hline 
specific\_data;	& The privacy policy declares the retention period or the criteria to determine it, but only for specific or limited types of data (e.g., social media comments data, etc.), but not for personal data rights in general or for dissected categories of personal data.	& 1 \\
\hline 
\rowcolor{gray!20}
\multicolumn{3}{l}{\textbf{Theme:} retention -> declaration\_type} \\
\hline 
\rowcolor{gray!20}
\textbf{Code}    & \textbf{Code Definition}   & \textbf{\#}     \\
\hline 
criteria;	& The privacy policy declares the retention period as one or more criteria (e.g., \enquote{as long as necessary to ...}). & 42 \\
\hline 
period; &	The privacy policy declares the retention period as one or more specific periods of time (e.g., \enquote{90 days})	& 8 \\
\hline 
\rowcolor{gray!20}
\multicolumn{3}{l}{\textbf{Theme:} retention -> data\_type} \\
\hline 
\rowcolor{gray!20}
\textbf{Code} [exclusive]   & \textbf{Code Definition}   & \textbf{\#}     \\
\hline 
multiple;	& The privacy policy declares the retention period for multiple data categories (e.g., account data, personalization data, marketing data, etc.) & 13 \\
\hline 
single;	& The privacy policy declares the retention period for a single data type or just states \enquote{personal data} or similar terms such as \enquote{personal information}, \enquote{personal details}, \enquote{information}, or \enquote{data} in general.	& 29 \\
\hline 
\rowcolor{gray!20}
\multicolumn{3}{l}{\textbf{Theme:} retention -> destruction\_method} \\
\hline 
\rowcolor{gray!20}
\textbf{Code}   & \textbf{Code Definition}   & \textbf{\#}     \\
\hline 
no;	[exclusive] & The privacy policy does not declare the method for destroying the data after the retention period is passed.	& 32 \\
\hline  
securely;	& The privacy policy declares that the data will be destroyed in a secure fashion after the retention period is passed.	& 10 \\
\hline 
anonymized;	& The privacy policy declares that the data will be anonymized so that re-identification is not possible after the retention period is passed.	& 2 \\
\hline 
promptly;	& The privacy policy declares that the data will be destroyed promptly after the retention period is passed.	& 6 \\
\bottomrule
\end{longtable}
\end{center}
}

{
\renewcommand{\arraystretch}{1.25}
\begin{center}
\small
\begin{longtable}{p{3cm}p{10.8cm}p{1.2cm}p{1cm}}
\caption{Codebook for the data subject rights to request the destruction and to obtain a copy of their personal data in a readable and clear format in the privacy policy text (survey Q\#38 -- Q\#47) (\textit{N = 91}). We use [destruct $|$ copy] in the table content when the code applies to both rights. Frequencies are presented separately for each right in the rightmost two columns. Themes are represented in grayed rows starting with \enquote{Theme}. The first column represents the code. Exclusive codes (i.e., cannot appear with other codes in the same theme are followed by \enquote{[exclusive]} either at the column label if this applies to all codes or at the code cell if it applies to a subset of the codes. The second column contains the code definition, and the last column contains the code frequency.} \label{tab:destruct_copy} \\
\endfirsthead
\caption[]{(continued)}\\
\endhead
\toprule 
\rowcolor{gray!20}
\multicolumn{4}{l}{\textbf{Theme:} [destruction | copy] -> declaration} \\
\hline 
\rowcolor{gray!20}
\textbf{Code} [exclusive]   & \textbf{Code Definition}   & \textbf{\protect\protect\makecell[l]{\#\\(destruct)}} & \textbf{\protect\makecell[l]{\#\\(copy)}}   \\
\hline 
yes;	& The privacy policy declares the right to request the [destruction | copy] of personal data.	& 46	& 31 \\ 
\hline 
no;	& The privacy policy does not declare the right to request the [destruction | copy] of personal data.	& 40		& 41 \\ 
\hline 
other;	& The privacy policy declares the right to request the [destruction | copy] of personal data applies only for specific cases (see the "other" sub-themes below), but not for personal data in general.	& 5 &	19 \\ 
\hline 
\rowcolor{gray!20}
\multicolumn{4}{l}{\textbf{Theme:}  [destruction | copy] -> declaration -> \enquote{other;} -> reason for \enquote{other;}} \\
\hline 
\rowcolor{gray!20}
\textbf{Code}   & \textbf{Code Definition}   & \textbf{\protect\makecell[l]{\#\\(destruct)}} & \textbf{\protect\makecell[l]{\#\\(copy)}}    \\
\hline 
access\_only; & The privacy policy declares the right to access data, but not to have a copy of it.	Unlike the GDPR, in the PDPL, the \enquote{right of access} (Art.5) is different from the \enquote{right to request access to personal data} (Art.6)~\cite{sdaia23_1}, where the latter grants data subjects the right to \enquote{request a copy of their Personal Data in a readable and clear format}~\cite{sdaia23_1}. & 0	& 19 \\ 
\hline 
account\_data\_only; & The privacy policy declares the right to the [destruction | copy | access] for account data only (or similar terms such as profile or registered on the site), but not personal data in general. & 4 &	0 \\
\hline 
conditional; & The privacy policy declares the right to the [destruction | copy | access] for account data only if a specific condition is met (e.g., if the user disagrees with policy changes).	& 1	 &	0 \\ 
\hline 
order\_data\_only;	& The privacy policy declares the right to the [destruction | copy | access] for purchase data only, but not personal data in general.	& 0	 &	1 \\ 
\hline 
\rowcolor{gray!20}
\multicolumn{4}{l}{\textbf{Theme:} [destruction | copy] -> method\_for\_exercising\_the\_right} \\
\hline 
\rowcolor{gray!20}
\textbf{Code}   & \textbf{Code Definition}   & \textbf{\protect\makecell[l]{\#\\(destruct)}} & \textbf{\protect\makecell[l]{\#\\(copy)}}    \\
\hline 
web\_form;	& The privacy policy declares a web form as a method for exercising the right to request the [destruction | copy] of personal data. This code also includes PDF forms.	& 8	 &  7 \\ 
\hline 
email;	& The privacy policy declares an email address to the controller as a method for exercising the right to request the [destruction | copy] of personal data. 	& 16	&	13 \\
\hline 
user\_account;	& The privacy policy declares adjusting the user account (or profile) settings (or preferences) as a method for exercising the right to request the [destruction | copy] of personal data.	& 11	& 	17 \\
\hline 
contact\_us;	& The privacy policy declares contacting the controller or checking the controller's contact details as a method for exercising the right to request the [destruction | copy] of personal data. The contact details can be inside the privacy policy text, or alternatively, they can be outside the privacy policy text (e.g., in the page's footer). See the sub-themes (starting with \enquote{contact}) of this theme in the next table. & 13	&	5 \\
\hline 
visit\_us;	& The privacy policy declares visiting the controller's physical location or checking the controller's address details as a method for exercising the right to request the [destruction | copy] of personal data. The physical location details can be inside the privacy policy text, or alternatively, they can be outside the privacy policy text (e.g., in the page's footer). See the sub-themes (starting with "visit") of this theme in the next table.	& 1	&	1 \\
\hline 
post\_mail;	& The privacy policy declares a post-mail address as a method for exercising the right to request the [destruction | copy] of personal data. 	& 2	 &	2 \\
\hline 
phone;	& The privacy declares a phone number as a method for exercising the right to request the [destruction | copy] of personal data. 	& 4	 &	2 \\ 
\hline 
no;	& The privacy policy does not declare the method for exercising the right to request the [destruction | copy] of personal data. This code also includes if the policy declares destruction method for a specific type of data (e.g., social media login details), but not personal data in general, as the question is about the method for exercising the right on personal data.	& 15	&	15 \\ 
\hline


\rowcolor{gray!20}
\multicolumn{4}{l}{\textbf{Theme:} [destruction | copy] -> method\_for\_exercising\_the\_righ -> [\enquote{contact us;} | \enquote{visit us;}] -> reference\_methods} \\
\hline 
\rowcolor{gray!20}
\textbf{Code}   & \textbf{Code Definition}   & \textbf{\protect\makecell[l]{\#\\(destruct)}} & \textbf{\protect\makecell[l]{\#\\(copy)}}    \\
\hline 
\protect\makecell[l]{contact\_footer\_email;}	& An email is provided in the footer area of the privacy policy (outside the policy text).	& 4	 & 1 \\ 
\hline 
\protect\makecell[l]{contact\_footer\_phone;}	& A phone number is provided in the footer area of the privacy policy (outside the policy text).	& 5	 &	1 \\ 
\hline 
\protect\makecell[l]{contact\_footer\_phys\_loc;}	& A physical location is provided in the footer area of the privacy policy (outside the policy text).	& 1	 &	0 \\ 
\hline 
\protect\makecell[l]{contact\_footer\_social\_\\media;}	& Social media accounts are provided in the footer area of the privacy policy (outside the policy text).	& 2	&	0 \\ 
\hline 
\protect\makecell[l]{contact\_footer\_whatsapp;}	& A WhatsApp link is provided in the footer area of the privacy policy (outside the policy text).	& 1	&	0 \\ 
\hline 
\protect\makecell[l]{contact\_footer\_website;}	& A link for the same website is provided in the footer area of the privacy policy (outside the policy text).	& 1	 &	0 \\ 
\hline 
\protect\makecell[l]{contact\_policy\_web\_form;}	& A web form link in the privacy policy's contact us area (inside the policy text). & 2	&	2 \\ 
\hline 
\protect\makecell[l]{contact\_policy\_email;}	& An email address in the privacy policy's contact us area (inside the policy text).	& 4	 &	2 \\ 
\hline 
\protect\makecell[l]{contact\_policy\_phone}	& A phone number in the privacy policy's contact us area (inside the policy text).	& 3	 &	1 \\ 
\hline 
\protect\makecell[l]{contact\_policy\_post\_mail;}	& A post-mail address in the privacy policy's contact us area (inside the policy text). &	2	&	3 \\ 
\hline 
\protect\makecell[l]{visit\_nearest\_store;}	& Refers data subjects to visit the physical location of the nearest store. &	1	&	1 \\ 
\hline 

\rowcolor{gray!20}
\multicolumn{4}{l}{\textbf{Theme:} [destruction | copy] -> response\_time} \\
\hline 
\rowcolor{gray!20}
\textbf{Code}    & \textbf{Code Definition}   & \textbf{\protect\makecell[l]{\#\\(destruct)}} & \textbf{\protect\makecell[l]{\#\\(copy)}}    \\
\hline 
general; & The privacy policy does not declare a specific time the controller takes to respond to [destruction | copy] of personal data requests. Instead, it declares that in general statements (e.g. \enquote{in time}, \enquote{in a reasonable time}, etc.) &	6 & 5 \\ 
\hline 
specific;	& The privacy policy declares a specific time the controller takes to respond to [destruction | copy] of personal data requests (e.g., 30 days). & 3	 &	2 \\ 
\hline 
no; [exclusive]  & The privacy policy does not declare how long the controller takes to respond to [destruction | copy] of personal data requests.	& 42	&	43 \\ 
\hline 

\rowcolor{gray!20}
\multicolumn{4}{l}{\textbf{Theme:} [destruction | copy] -> free\_of\_charge} \\
\hline 
\rowcolor{gray!20}
\textbf{Code} [exclusive]    & \textbf{Code Definition}   & \textbf{\protect\makecell[l]{\#\\(destruct)}} & \textbf{\protect\makecell[l]{\#\\(copy)}}    \\
\hline 
yes;	& The privacy policy declares that processing data [destruction | copy] requests is free of charge.  & 6	&	6 \\ 
\hline 
no;	& The privacy policy declares that processing the data [destruction | copy] requests is not free of charge (paid).	& 0	 &	0 \\ 
\hline 
n.m.	& The privacy policy does not declare that processing data [destruction | copy] requests is free of charge. This does not imply it is paid, but just not mentioned.	& 45	&	44 \\ 
\hline 
\rowcolor{gray!20}
\multicolumn{4}{l}{\textbf{Theme:} copy -> clear\_and\_readable\_copy} \\
\hline 
\rowcolor{gray!20}
\textbf{Code}   & \textbf{Code Definition}   & \textbf{\protect\makecell[l]{\#\\(destruct)}} & \textbf{\protect\makecell[l]{\#\\(copy)}}    \\

yes; [exclusive] 	 & The privacy policy's declaration for the right to obtain a copy of personal data includes the statement \enquote{in a clear and readable format}.	& n.a.	&	3 \\ 
\hline 

no; [exclusive] 	& The privacy policy's declaration for the right to obtain a copy of personal data does not include the statement \enquote{in a clear and readable format}.	& n.a.	&	33 \\ 
\hline 

yes\_indirect; [exclusive] 	 & The privacy policy's declaration for the right to obtain a copy of personal data includes an indirect statement indicating clear and readable format, such as \enquote{usable electronic format} and \enquote{customary format}, \enquote{standard, readable format},  \enquote{structured, machine-readable format} (without portability right -- if portability right is declared explicitly, this has a separate code).	& n.a.	&	5 \\ 
\hline 

portability;  &	The privacy policy contains a declaration for the portability right, whether with or without the declaration for the right to obtain a copy. It often includes \enquote{machine-readable format}, \enquote{structured, machine-readable format}.  &	n.a.	&	9 \\ 

\bottomrule
\end{longtable}
\end{center}
}

{
\renewcommand{\arraystretch}{1.25}
\begin{center}
\small
\begin{longtable}{p{3.5cm}p{12cm}p{1cm}}
\caption{Codebook for the mechanism to file complaints related to data rights and
processing in the privacy policy text (survey Q\#53 -- Q\#57) (\textit{N = 91}). Themes are represented in grayed rows starting with \enquote{Theme}. The first column represents the code. Exclusive codes (i.e., cannot appear with other codes in the same theme are followed by \enquote{[exclusive]} either at the column label if this applies to all codes or at the code cell if it applies to a subset of the codes. The second column contains the code definition, and the last column contains the code frequency.} \label{tab:complaint} \\
\endfirsthead
\caption[]{(continued)}\\
\endhead
\toprule 
\rowcolor{gray!20}
\multicolumn{3}{l}{\textbf{Theme:} complaint -> declaration} \\
\hline 
\rowcolor{gray!20}
\textbf{Code} [exclusive]    & \textbf{Code Definition}   & \textbf{\protect\makecell[l]{\#}}  \\
\hline 
yes;	& The privacy policy declares a mechanism for filing complaints.	& 30 \\ 
\hline 
no;	& The privacy policy does not declare a mechanism for filing complaints.	& 55 \\ 
\hline 
other;	& The privacy policy declares a mechanism for filing complaints related to data rights and processing, but in a format different from what is expected (see the \enquote{other} sub-themes below).	& 6 \\ 
\hline 
\rowcolor{gray!20}
\multicolumn{3}{l}{\textbf{Theme:} complaint -> declaration -> "other;" -> reason for "other;"} \\
\hline 
\rowcolor{gray!20}
\textbf{Code}   & \textbf{Code Definition}   & \textbf{\protect\makecell[l]{\#}}  \\
\hline 
right\_to\_object;	& The privacy policy declares the \enquote{right to object} (which is normally in the rights section), but no separate section for complaints. & 6  \\
\hline 

\rowcolor{gray!20}
\multicolumn{3}{l}{\textbf{Theme:} complaint -> separate\_section} \\
\hline 
\rowcolor{gray!20}
\textbf{Code} [exclusive]    & \textbf{Code Definition}   & \textbf{\protect\makecell[l]{\#}}  \\
\hline 
yes;	& The privacy policy has a dedicated section for complaints.	& 17 \\ 
\hline 
no;	& The privacy policy does not have a dedicated section for complaints.	& 19 \\
\hline 

\rowcolor{gray!20}
\multicolumn{3}{l}{\textbf{Theme:} complaint -> internal\_dept\_officer\_name} \\
\hline 
\rowcolor{gray!20}
\textbf{Code} [exclusive]    & \textbf{Code Definition}   & \textbf{\#}  \\
\hline 
yes;	& The privacy policy declares the department and/or officer name responsible for complaints related to personal data.	& 12 \\ 
\hline 
no;	& The privacy policy does not declare the department and/or office name responsible for complaints related to personal data.	& 24 \\ 
\hline 

\rowcolor{gray!20}
\multicolumn{3}{l}{\textbf{Theme:} complaint -> channels} \\
\hline 
\rowcolor{gray!20}
\textbf{Code}   & \textbf{Code Definition}   & \textbf{\protect\makecell[l]{\#}}  \\
\hline 
email;	& The privacy policy declares an email address as a channel for complaints related to personal data.	& 22 \\ 
\hline 
post\_mail;	& The privacy policy declares a post-mail address as a channel for complaints related to personal data.	& 6 \\
\hline 
phys\_loc;	& The privacy policy declares visiting a physical location address as a channel for complaints related to personal data.	& 1 \\
\hline 
website;	& The privacy declares the website (the word \enquote{website} or puts the website's address) as a channel for complaints related to personal data.	& 1 \\
\hline 
contact\_us;	& The privacy policy declares that contacting the controller or checking the controller's contact details as a channel for complaints related to personal data. The contact details can be inside the privacy policy text, or alternatively, they can be outside the privacy policy text (e.g., in the page's footer). See the sub-themes of this theme in the next table.	& 3 \\ 
\hline 
visit\_us;	& The privacy policy declares visiting the controller's physical location or checking the controller's address details as a channel for complaints related to personal data. The physical location details can be included in the privacy policy text or placed elsewhere (e.g., in the page footer). See the sub-themes of this theme in the next table. &  2 \\
\hline 
\protect\makecell[l]{contact\_auth\_general;}	& The privacy policy declares contact authority as a first channel for complaints related to personal data. The provided authority name is general (e.g., your local authority).	& 4 \\ 
\hline 
\protect\makecell[l]{contact\_\\auth\_specific;}	& The privacy policy declares contact authority as a first channel for complaints related to personal data. The provided authority name is specific (e.g., SDAIA).	& 1 \\
\hline 
user\_account;	& The privacy policy declares adjusting the user account (or profile) settings (or preferences) as a channel for complaints related to personal data.	& 1 \\
\hline 
no;	& The privacy policy does not declare a channel for complaints related to personal data. 	& 2 \\ 

\rowcolor{gray!20}
\multicolumn{3}{l}{\textbf{Theme:} complaint -> channel -> [\enquote{contact us} | \enquote{visit us}] -> reference\_channels} \\
\hline 
\rowcolor{gray!20}
\textbf{Code}   & \textbf{Code Definition}   & \textbf{\protect\makecell[l]{\#}}  \\
\hline 
\protect\makecell[l]{contact\_footer\_whatsapp;} & A WhatsApp link is provided in the footer area of the privacy policy (outside the policy text).	& 1 \\
\hline 
\protect\makecell[l]{contact\_footer\_phone;}	& A phone number is provided in the footer area of the privacy policy (outside the policy text).	& 1 \\
\hline 
\protect\makecell[l]{contact\_footer\_post\_mail;} &	A post-mail address is provided in the footer area of the privacy policy (outside the policy text).	& 1 \\ 
\hline 
\protect\makecell[l]{contact\_policy\_post\_mail;}	& A post-mail address is provided in the privacy policy's contact us area (inside the policy text).	& 1 \\ 
\hline 
\protect\makecell[l]{contact\_policy\_web\_form;}	& A web form link is provided in the privacy policy's contact us area (inside the policy text).	& 1 \\ 
\hline 
\rowcolor{gray!20}
\multicolumn{3}{l}{\textbf{Theme:} complaint -> response\_time} \\
\hline 
\rowcolor{gray!20}
\textbf{Code} [exclusive]   & \textbf{Code Definition}   & \textbf{\protect\makecell[l]{\#}}  \\
\hline 
general;	& The privacy policy does not declare a specific time the controller takes to respond to complaints related to personal data requests. Instead, it declares that in general statements (e.g., \enquote{in time}, \enquote{in a reasonable time}, etc.) & 3 \\
\hline 
specific;	& The privacy policy declares a specific time the controller takes to respond to complaints related to personal data requests (e.g., 30 days). & 1 \\ 
\hline 
no;	 & The privacy policy does not declare how long the controller takes to respond to complaints related to personal data & 32 \\ 
\hline 
\rowcolor{gray!20}
\multicolumn{3}{l}{\textbf{Theme:} complaint -> free\_of\_charge} \\
\hline 
\rowcolor{gray!20}
\textbf{Code} [exclusive]   & \textbf{Code Definition}   & \textbf{\protect\makecell[l]{\#}}  \\
\hline 
yes;	& The privacy policy declares that processing complaints is free of charge. 	& 1 \\
\hline 
no;	& The privacy policy declares that processing complaints is not free of charge (paid).	& 0 \\ 
\hline 
n.m.	& The privacy policy does not declare that processing complaints is free of charge. This does not imply it is paid, but just not mentioned.	& 99 \\ 
\hline 
\rowcolor{gray!20}
\multicolumn{3}{l}{\textbf{Theme:} complaint -> authority\_as\_second\_resolution} \\
\hline 
\rowcolor{gray!20}
\textbf{Code}    & \textbf{Code Definition}   & \textbf{\protect\makecell[l]{\#}}  \\
\hline 

general; [exclusive]	& The privacy policy declares a general name for data authority as a second solution channel that the subject can raise the complain to (e.g., your local data authority), if the outcome of the complaint is not satisfactory.	& 7 \\ 
\hline 

specific; [exclusive]	& The privacy policy declares a specific name for data authority as a second solution channel that the subject can raise the complain to (e.g., SDAIA), if the outcome of the complaint is not satisfactory.	& 6 \\ 
\hline 
auth\_outside\_ksa;	& The declared data authority (whether general or specific) is outside KSA.	& 3 \\ 
\hline 
no;	[exclusive] & The privacy policy does not declare either a general or specific data authority as a second solution channel that the subject can raise a complaint to, if the outcome of the initial complaint is not satisfactory.	& 23 \\ 
\bottomrule
\end{longtable}
\end{center}
}

\renewcommand{\arraystretch}{1.25}
\begin{table*}[!h]
\small
\caption{Codebook for potential causes of discrepancies between human and LLM privacy policy answers. The first column represents the main codes, which are exclusive (cannot appear twice), followed by their definitions and frequencies columns. The fourth column contains the subcodes, which are non-exclusive, followed by their definitions and frequencies columns. The } 
\label{tab:themes}
\begin{tabular}{p{3cm}p{3cm}p{0.5cm}p{2.5cm}p{6cm}p{0.5cm}}
\toprule 
\rowcolor{gray!20}
\multicolumn{6}{l}{\textbf{Theme:} human\_vs\_llm\_discrepancy\_reason} \\
\hline 
\rowcolor{gray!20}
\makecell[l]{\textbf{Code}\\\textbf{[exclusive]}}  & \textbf{Definition} & \#  & \makecell[l]{\textbf{Subcode}\\\textbf{[non-exclusive]}}    & \textbf{Definition}  &  \textbf{\#}  \\
\hline 
\multirow{5}{3cm}{human\_error;}	& \multirow{5}{3cm}{The discrepancy is due to human error.} & \multirow{5}{*}{4} & complex\_stmnt;	& The privacy policy contains complex statements such as conditional statements. & 1 \\ 
\cline{4-6}
 & & & no\_specific\_section;  & The privacy policy contains statements not under a specific section, e.g., under the main title \enquote{privacy policy}. & 3 \\ 
\cline{4-6}
 & & & long\_policy; & The privacy policy is overlay long. & 1 \\ 
\cline{4-6}
 & & & accordion\_menu; &	The privacy policy contains accordion menus that hinder the browser's search functionality. & 1 \\ 
\cline{4-6}
 & & & scattered; & The privacy policy scatters a specific right declaration among several sections, e.g., retention statements divided into many sections. & 1 \\
\hline 
llm\_error;	& The discrepancy is due to LLM error. & 2 & outside\_policy\_text;  &	The LLM analyzed text outside the privacy policy text boundaries (e.g., page's footer). & 2 \\
\hline 
llm\_more\_restrictive;	& The LLM answer is more restrictive than that of the human. That is, LLM answered \enquote{no} or \enquote{other} for a human's \enquote{yes}, or answered \enquote{no} for a human's \enquote{other}. & 40 & literal\_interpretation &	The LLM interpreted the question literally, ignoring indirect soft statements that are commonly used to convey the same meaning. & 40 \\
\hline 
\multirow{4}{3cm}{llm\_more\_permissive;}	& \multirow{4}{3cm}{The LLM answer is more permissive than that of the human. That is, it answered \enquote{yes} or \enquote{other} for a human's \enquote{no}, or answered \enquote{yes} for a human's \enquote{other}.} & \multirow{4}{*}{20} & over\_analysis;	& The LLM overanalyzed statements in the policy, leading to inaccurate or incorrect conclusions that are not explicitly stated in the policy. & 2  \\
\cline{4-6}
	& & & misaligned\_context;	& The LLM answer was drawn from a privacy policy with a context different than the human's, such as different laws or terminology from another country. & 10 \\ 
\cline{4-6}
 & & & generalization;	& The LLM answer generalized a limited statement, such as a right declared on specific data, not all data. & 5 \\ 
\cline{4-6}
& & & literal\_interpretation	& The LLM interpreted the question literally, ignoring the meaning of the context. For example, stating the policy declared a right just because there was a heading mentioning a word about the right without any meaningful text. & 3 \\
\bottomrule
\end{tabular}
\end{table*}

\end{document}